\documentclass[12pt]{iopart}
\usepackage{iopams}
  
\usepackage{epsfig}  

\newcommand{\beq}{\begin{equation}}
\newcommand{\eeq}{\end{equation}}
\newcommand{\bea}{\vspace{0.25cm}\begin{eqnarray}}
\newcommand{\eea}{\end{eqnarray}}

\newcommand{\ro}{\mbox{{\boldmath
$\rho$}}}

\newcommand{\qb}{\mbox{{\bf
q}}}
\newcommand{\pb}{{{\bf p}}}

\newcommand{\bb}{{{\bf b}}}

\newcommand{\vb}{{{\bf v}}}

\def\lsim{\mathrel{\rlap{\lower4pt\hbox{\hskip1pt$\sim$}}
    \raise1pt\hbox{$<$}}}         
\def\gsim{\mathrel{\rlap{\lower4pt\hbox{\hskip1pt$\sim$}}
    \raise1pt\hbox{$>$}}}         

\begin{document}

\title[]{Updated analysis of jet quenching at RHIC and LHC within the
light cone path integral approach
}

\author{B.G. Zakharov}

\address{
L.D.~Landau Institute for Theoretical Physics,
        GSP-1, 117940,\\ Kosygina Str. 2, 117334 Moscow, Russia
}

\ead{bgz@itp.ac.ru}
\begin{abstract}
  We present results of a detailed analysis of experimental
  data   on the nuclear modification factor $R_{AA}$ and the flow coefficient
  $v_2$ for light hadrons
  from RHIC for $0.2$ TeV Au+Au collisions and from LHC
  for $2.76$ and $5.02$ TeV Pb+Pb, and $5.44$ TeV
  Xe+Xe collisions.
  We perform calculations within the light-cone path
  integral approach to induced gluon emission.
We use running $\alpha_s$ which is frozen
  at low momenta at some value $\alpha_{s}^{fr}$.
  We find that the RHIC data support somewhat larger value of $\alpha_{s}^{fr}$.
  For the $\chi^2$ optimized values of $\alpha_{s}^{fr}$, the theoretical
  predictions are in reasonable agreement with data on $R_{AA}$
  and $v_2$.  
  Calculations made for different formation times and life-times
  of the QGP show that
jet quenching at the RHIC and LHC energies
   is only weakly sensitive to the initial and final stages
  of the QCD matter evolution.
\end{abstract}





\section{Introduction}
It is widely believed that strong suppression of the high-$p_T$
particle spectra in $AA$ collisions (usually called the jet quenching)
observed first at RHIC, and later at the LHC,
is due to
parton energy loss in the quark-gluon plasma (QGP)
produced in the initial stage of nucleus collisions
(for reviews on the jet quenching phenomenon, see, e.g., \cite{JQ1,JQ2}).
For the RHIC and LHC conditions, 
the dominating contribution to the parton energy loss comes from
the induced gluon radiation caused by parton multiple scattering in the
QGP \cite{GW,BDMPS1,BDMPS_pt,LCPI1,W1,GLV1,AMY1}, and
the effect of the collisional energy loss \cite{Bjorken1}
turns out to be relatively weak \cite{BSZ,Z_coll,Gale_coll}.
In the pQCD picture, for the QGP modeled by a system of the Debye
screened color centers
\cite{GW},
the induced gluon spectrum can be expressed  via  the Green function  of
a  2D Schr\"odinger equation with an imaginary potential
\cite{LCPI1,BDMPS1}, in which the longitudinal coordinate $z$ plays the role
of time.
This potential is $\propto n\sigma_{q\bar{q}}(\rho)$, where
$n$ is the QGP number density and $\sigma_{q\bar{q}}(\rho)$ is
the dipole cross section  for scattering of
a color singlet $q\bar{q}$ pair off the QGP constituent (here, 
$\rho$ is the size of the $q\bar{q}$-pair).
For the  quadratic approximation $\sigma_{q\bar{q}}(\rho)\approx C\rho^2$,
the Hamiltonian of the Schr\"odinger equation takes the harmonic
oscillator (HO) form
with  a  complex  frequency $\Omega^2\propto \hat{q}$ with $\hat{q}=2Cn$.
At the same time, the quantity $\hat{q}$, commonly called the transport
coefficient, characterizes the $L$-dependences of the parton
$p_T$-broadening in the medium:
$\langle p_T^{2}\rangle=L\hat{q}$ \cite{BDMPS_pt}.

In the HO approximation the induced gluon spectrum
for massless partons in a uniform medium 
can be evaluated analytically \cite{BDMPS1,BDMS98}. In
\cite{BDMS_dynscal} it was shown that, for the Bjorken
like QGP expansion \cite{Bjorken}, the total radiative energy loss in the HO
approximation can be expressed via that for a static medium with
an equivalent linear averaged transport coefficient: $\hat{q}_{st}=
\frac{2}{L^2}\int d\tau \tau \hat{q}(\tau)$.
By numerical calculations it was found \cite{WS_dynscal}
that such a dynamical scaling, to rather good accuracy, holds also
for the gluon spectrum.
Making use of this dynamical scaling law simplifies greatly jet quenching
calculations for an expanding QGP \cite{Eskola}.
However, in a recent analysis \cite{Salgado_dynscal} it was demonstrated
that the approximation of the
dynamical scaling may be too crude 
for precise modeling of the jet quenching phenomenon.
 In any case, the HO approximation itself cannot be regarded
  as satisfactory for accurate jet quenching simulations.
{  The apparent shortcoming of the HO approximation is that the opacity
  expansion series of the gluon spectrum for massless partons  does
  not contain the $N=1$
(and all odd terms) rescattering contribution \cite{Z_OA}
(see also \cite{Arnold_OA}). 
The $N=2k+1$ rescattering terms become nonzero
due to the Coulomb effects, that lead
to logarithmic dependence of the factor $C=\sigma_{q\bar{q}}(\rho)/\rho^2$
at $\rho\to 0$, and mass effects \cite{Z_OA,AZ}. These effects may
change drastically the HO gluon spectrum in the
regime when gluon formation length
becomes comparable or larger than the parton path
length in the QGP \cite{Z_OA,AZ}.}

The absence of the leading $N=1$ term in the HO approximation
is not very important in the limit of strong Landau-Pomeranchuk-Migdal
(LPM) suppression, when the typical number of rescatterings becomes very large.
However, for the QGP produced in $AA$-collisions we have a situation
when the contribution of the $N=1$ term dominates the radiative
parton energy loss \cite{GLV1}.
This fact stimulated the jet quenching analyses
(see, e.g., \cite{CUJET3,Djord1} and references therein)
based on the GLV formalism
\cite{GLV1} with accounting for only $N=1$ term. 
In the models restricted to the $N=1$ rescattering, 
the effect of higher order rescatterings can be partly absorbed
into a redefinition of $\alpha_s$ (or some of the QGP parameters).
But this should influence the predictions
for quantities that depend on variation of
the LPM suppression (because the magnitude of the LPM effect varies
significantly with the parton energy, the QGP size and density).
As a result, such important jet quenching characteristics as the $p_T$- and
centrality-dependence of the $R_{AA}$, azimuthal flow coefficient $v_2$
may have considerable theoretical uncertainties. 

{ In this work we continue our investigations
  \cite{RAA08,RAA11,RAA11JP,RAA13}
of the jet quenching within the light-cone path integral (LCPI)
approach \cite{LCPI1} to the
induced gluon emission.
The main motivation of the present work is to
perform a more comprehensive analysis of the data from RHIC and LHC
including the recent LHC data for $5.02$ TeV
Pb+Pb and $5.44$ TeV  Xe+Xe collisions.}
The LCPI formalism \cite{LCPI1}
is free from the above mentioned problems inherent
to the models based on the HO and $N=1$ approximations.
It allows one to perform calculations for an arbitrary number of rescatterings
beyond the HO approximation for massive partons with accurate treatment
of the finite-size effects.
Also, it is free from the restriction to the strong
LPM suppression (which is needed in the BDMPS formulation \cite{BDMPS1}).
The LCPI formalism \cite{LCPI1} is applicable to
any induced process of the type $a\to bc$ both in QED and QCD.
It is based on the path integral representation in the coordinate space
for the in-medium wave functions of fast particle
on the light-cone $t-z=$const.
For the case of a uniform, infinite medium the LCPI formulation
is equivalent \cite{AZ_AMY} to the the AMY \cite{AMY1} approach,
formulated in the momentum space.
The induced gluon/photon spectrum
was originally written in \cite{LCPI1} in terms of Green's function
of the Schr\"odinger equation. Since the Green function is singular
at $z\to 0$, the Green function representation is inconvenient for numerical
calculations beyond the HO approximation.
In \cite{RAA04}, by solving the Schr\"odinger equation backward in
time/$z$, we obtained a representation
of the induced gluon spectrum for a finite-size medium in terms
of a solution to the Schr\"odinger equation with a smooth boundary
condition, which is convenient for numerical computations (this
form has been rediscovered later in \cite{Gale-Caron}).

Unfortunately, numerical solving the Schr\"odinger equation
requests a rather large computational time. 
This renders difficult numerical jet quenching simulations 
for realistic scenarios of the QGP formation and evolution.
This forces to use simplified models for the initial QGP fireball
and its hydrodynamic evolution, in which
the computational cost may be reduced.
In \cite{RAA08}
we have performed calculations of
the nuclear modification factor $R_{AA}$
within the LCPI scheme \cite{RAA04} with realistic dipole cross section
using the model of the QGP fireball
with a uniform density distribution in the transverse plane and
with the Bjorken 1+1D hydrodynamical longitudinal expansion.
For this model the density profile along the fast parton trajectory
for each jet production point turns out to be the same (only its length $L$
varies).
This makes it possible to perform first tabulating the $L$-dependence of
the induced gluon spectrum and then to use interpolation
in computation of the medium modified jet fragmentation functions (FFs)
for arbitrary positions of the jet production. This procedure
greatly reduces the computational time.
In \cite{RAA11,RAA11JP,RAA13} we used the model \cite{RAA08} 
for analysis of the RHIC and LHC data on the nuclear modification factor
$R_{AA}$ in $0.2$ TeV Au+Au collisions at RHIC
and $2.76$ TeV Pb+Pb collisions at the LHC. 
It was found that  predictions of the LCPI approach are in reasonable
agreement with data. But this is only possible if one uses somewhat larger
$\alpha_s$ for RHIC. 

In the present work, as in \cite{RAA11,RAA11JP,RAA13} we use the method of
\cite{RAA04} for calculations of the induced gluon spectrum
with realistic dipole cross section. We use a somewhat improved
version of the scheme of \cite{RAA08} for evaluating { the medium modified
FFs from the one gluon spectrum (see appendix B for details).} 
We perform comparison with data on both the nuclear modification factor
$R_{AA}$ and the azimuthal anisotropy $v_2$ for different centralities.
Contrary to our previous jet quenching analyses
\cite{RAA11,RAA11JP,RAA13}, where a visual comparison with data
was performed, in the present study we perform an accurate $\chi^2$ fit
to the data. Our results for the nuclear modification factor $R_{AA}$
agree quite well with experimental data. Although we do not
include the data on $v_2$ in the $\chi^2$ analysis, our
predictions for $v_2$ are in reasonable agreement
with the data as well. However, similarly to our previous
analyses, we find that the description of the RHIC data requires
somewhat bigger $\alpha_s$
than that for the LHC. In the present study it is demonstrated with the help
of an accurate $\chi^2$ analysis.

The paper is organized as follows. In Section 2, we
briefly review
the basic aspects of our jet quenching model.
In  Section  3, we perform
the $\chi^2$ fit of the experimental data on $R_{AA}$
for finding the optimal $\alpha_s$, 
and then confront the theoretical
results with data on $R_{AA}$ and $v_2$.
Conclusions  are  contained  in Section 4. 
Some of the details of our calculations are given in two appendices.
In appendix A, we give the basic formulas for evaluation of the induced
gluon spectrum. The method for calculation of the in-medium FFs is
considered in appendix B.

\section{Theoretical framework}
In this section  we outline the main aspects of our method
for calculating the nuclear modification factor $R_{AA}$
and the azimuthal anisotropy $v_2$. It is similar to the method
used in our previous jet quenching analyses \cite{RAA08,RAA11,RAA13},
to which the interested reader is referred for details.

\subsection{Nuclear modification factor}
We will consider the central rapidity region around $y=0$.
We write the nuclear modification factor $R_{AA}$ 
for given impact parameter $b$ of $AA$-collision,
the hadron transverse momentum $\pb_T$
and rapidity $y$
as
\beq
R_{AA}(b,\pb_{T},y)=\frac{{dN(A+A\rightarrow h+X)}/{d\pb_{T}dy}}
{T_{AA}(b){d\sigma(N+N\rightarrow h+X)}/{d\pb_{T}dy}}\,,
\label{eq:10}
\eeq
where $T_{AA}(b)=\int d\ro T_{A}(\ro) T_{A}(\ro-\bb)$,
$T_{A}(\ro)=\int dz \rho_A(\sqrt{\rho^2+z^2})$ is the nuclear thickness function
(with $\rho_A$ the nuclear density).
The nominator of (\ref{eq:10})
is the differential yield of the 
process $A+A\to h+X$ (for clarity we omit the arguments $b$ and $\pb_T$).
It can be written in terms of the medium-modified hard cross section
$d\sigma_{m}/d\pb_{T} dy$ as
\beq
\hspace{-2cm}\frac{dN(A+A\rightarrow h+X)}{d\pb_{T} dy}=\int d\ro T_{A}(\ro+\bb/2)
T_{A}(\ro-\bb/2)
\frac{d\sigma_{m}(N+N\rightarrow h+X)}{d\pb_{T} dy}\,.\,\,\,
\label{eq:20}
\eeq
We write $d\sigma_{m}/d\pb_{T} dy$ in the form 
\beq
\frac{d\sigma_{m}(N+N\rightarrow h+X)}{d\pb_{T} dy}=
\sum_{i}\int_{0}^{1} \frac{dz}{z^{2}}
D_{h/i}^{m}(z, Q)
\frac{d\sigma(N+N\rightarrow i+X)}{d\pb_{T}^{i} dy}\,,\,\,\,
\label{eq:30}
\eeq
where $\pb_{T}^{i}=\pb_{T}/z$ is the transverse momentum
of the initial hard parton, 
${d\sigma(N+N\rightarrow i+X)}/{d\pb_{T}^{i} dy}$ is the
ordinary hard cross section,  
$D_{h/i}^{m}(z,Q)$ is the medium-modified FF
for transition of a parton $i$ with the virtuality $Q\sim p^i_T$ to the
final particle $h$.
The FF $D_{h/i}^{m}$ depends on the medium parameters along the
jet path in the medium:
\beq
\ro(\tau)=\ro_j+\vb\tau\,,
\label{eq:40}
\eeq
where $\ro_j$ is transverse coordinate of the hard process, and
$\vb\approx \pb_T/|\pb_T|$ is
the jet velocity. Since $|\vb|\approx 1$, the proper time  
$\tau$ in (\ref{eq:40}) coincides with the jet path length.
The variation of the medium profile with the jet azimuthal angle
generates the dependence of the nuclear modification factor
on the direction of the hadron momentum relative to the reaction
plane. We consider the event-averaged smooth evolution of the QGP fireball.
In this approximation, the azimuthal dependence can be characterized by the even
azimuthal coefficients $v_{2n}$, which, for a given impact parameter
can be written as
\beq
v_{2n}(p_\perp)=\frac{1}{2\pi R_{AA}(b,p_\perp)}
\int d\phi R_{AA}(b,\pb_{T},y)\cos(2n\phi)\,,
\label{eq:50}
\eeq
where
\beq
R_{AA}(b,p_\perp)=\frac{1}{2\pi}
\int d\phi R_{AA}(b,\pb_{T},y)
\label{eq:60}
\eeq
is the azimuthally averaged nuclear modification factor.
For comparison with experimental quantities measured in
a centrality bin $\Delta (c_1,c_2)$, in the above formulas one should perform
integration over the impact parameter region $(b_1,b_2)$ with $b_{1,2}$
written in terms of the $c_{1,2}$ (the $b$-integrations should be
performed separately for the numerators and denominators).
Experimentally, the centrality of an event
is defined in terms of the soft charged particle multiplicity
(in some kinematic region).
To a very good accuracy (except for the  most peripheral collisions)
the relation between the centrality and the impact parameter reads
$c=\pi b^2/\sigma_{in}^{AA}$ \cite{centrality}.

As usual, we assume that the final particles are formed outside the medium.
We also assume that for medium-modified parton-to-parton FFs,
the DGLAP stage precedes the induced gluon emission stage.
With the help of the formation length arguments, one can show that
this approximation should be reasonable at least for jets with
energy $\lsim 100$ GeV \cite{RAA08}.
In this approximation, $D_{h/i}^{m}$ 
symbolically
can be written as
\beq
D_{h/i}^{m}(Q)\approx D_{h/j}(Q_{0})
\otimes D_{j/k}^{in}\otimes D_{k/i}(Q)\,,
\label{eq:70}
\eeq
where $\otimes$ denotes $z$-convolution, 
$D_{k/i}$ is the DGLAP parton FF for $i\to k$ transition,
$D_{j/k}^{in}$ is the in-medium FF for $j\to k$ transition
due to induced gluon emission, and 
$D_{h/j}$ is the vacuum FF for transition of the parton $j$ to
the final particle $h$.

For the vacuum FFs ${D}_{h/j}(z,Q_0)$ we use the KKP
\cite{KKP} parametrization with $Q_0=2$ GeV.
In numerical calculations we compute the DGLAP FFs
$D_{k/i}$ by interpolating from a 2D $(z,Q)$ grid, created
with the help of the PYTHIA event generator \cite{PYTHIA}.
The FFs for $pp$ collisions needed to calculate $R_{AA}$ have
been obtained by the convolution of the KKP FFs at $Q_0=2$ GeV and the DGLAP
FFs. This method for the $pp$ FFs reproduces reasonably well the $Q$-dependence 
of the KKP FFs \cite{KKP}. However, the procedure with the same DGLAP
FFs for $AA$ and $pp$ collisions is preferable since
it guarantees that in the limit of
vanishing induced radiation, the medium-modified FFs $D_{h/i}^{m}$
exactly reduce to the $pp$ FFs.
We calculate the FFs $D_{j/k}^{in}$ from the one gluon induced spectrum
accounting for multiple gluon emission in the approximation
of independent gluon radiation \cite{BDMS_RAA}. The formulas for calculation
of the gluon induced spectrum and the FFs $D_{j/k}^{in}$ are given
in appendices A and B.
Note that in numerical calculations, instead of performing the convolution
of the FFs, as in (\ref{eq:70}), to use the
full FFs $D_{h/i}^{m}$ in (\ref{eq:30}),
we calculated sequentially the cross sections with the help of
formula (\ref{eq:30}) for the FFs in each of stage. This is technically
more convenient because at each stage we deal with the cross sections
which (in logarithmic scales)  are smooth functions.

As in \cite{RAA08,RAA13},   we account for
the effect of the collisional parton energy loss (which
is relatively small \cite{Z_coll,Gale_coll})
by redefining, in calculating the FFs $D_{j/k}^{in}$, the initial QGP 
temperature (for each geometry of the jet production) according to the
condition
\beq
\Delta E_{rad}(T^{\,'}_{0})=\Delta E_{rad}(T_{0})+\Delta E_{col}(T_{0})\,,
\label{eq:80}
\eeq
where
$\Delta E_{rad/col}$ is the radiative/collisional energy loss, $T_{0}$
is the real initial temperature of the QGP, and $T^{\,'}_{0}$ is the 
renormalized temperature.
We calculate $\Delta E_{col}$ using the Bjorken method \cite{Bjorken1}
with an accurate treatment of kinematics of the binary collisions
(the details can be found \cite{Z_coll}). The effect of collisional
energy loss on jet quenching is relatively small.
For RHIC the collisional mechanism
reduces $R_{AA}$ by $\sim 15-20$\% at $p_T\sim 10-20$ GeV, and
for the LHC energies the reduction of $R_{AA}$ is $\sim 5-15$\%
for $p_T\sim 10-100$ GeV (the effect becomes smaller at higher $p_T$).

We evaluate the induced gluon spectrum for the QGP modeled by a 
system of the static Debye  screened color centers \cite{GW}.
We assume that the number density of the QGP constituents
can be obtained from the entropy in the ideal gas model.
The assumption that $n\propto s$
seems to be reasonable, but has no theoretical justification.
We perform calculations using the Debye mass obtained in
the lattice analysis \cite{Bielefeld_Md}, 
which gives $\mu_{D}/T$ slowly decreasing with $T$  
($\mu_{D}/T\approx 3.2$ at $T\sim T_{c}$, $\mu_{D}/T\approx 2.4$ at 
$T\sim 4T_{c}$).
For the $T$-dependence of the Debye mass
we use the temperature extracted from the
lattice entropy density obtained in \cite{t-lat}. For a given entropy,
this temperature is somewhat larger than the ideal gas temperature (see below).
However, for the Debye mass defined via the ideal gas temperature
the results do not change significantly.
In our calculations of the induced gluon spectrum,
we take $m_{q}=300$ and $m_{g}=400$ MeV 
for the light quark and gluon quasiparticle masses
in the QGP. These values are
supported by the analysis of the lattice data  within the quasiparticle model
\cite{LH}. Note that the results are not very sensitive to the gluon
mass, and are practically insensitive to the light quark mass.

We perform numerical calculations of the nuclear modification 
using for the hard cross sections
on the right-hand side of
(\ref{eq:30}) the LO  pQCD formula with the CTEQ6 \cite{CTEQ6}
parton distribution functions.
In calculating the hard cross sections, we use
for the virtuality scale in $\alpha_{s}$ the value 
$cQ$ with $c=0.265$ as in the PYTHIA event generator \cite{PYTHIA},
which allows to simulate the higher order effects, and 
gives a fairly good description of the high-$p_{T}$ spectra in
$pp$-collisions. Note however that for nuclear modification
factor a moderate variation of the hard partonic cross sections is not
very crucial.
We account for the nuclear modification of the parton distributions
with the EPS09 correction \cite{EPS09} (we have found that the
difference in the results for EPS09 \cite{EPS09} and EKS98 \cite{EKS98}
corrections is visually negligible).
In calculating the distribution of the jet production points in
the transverse plane in formula (\ref{eq:20})
and the overlap function $T_{AA}$ in (\ref{eq:10}),
we use the Woods-Saxon nuclear density
$\rho_{A}(r)=\rho_{0}/[1+\exp((r-R_{A})/d)]$. For Au and Xe nuclei we take
$d=0.54$ fm and  $R_{A}=(1.12A^{1/3}-0.86/A^{1/3})$ fm
as in the GLISSANDO Glauber model \cite{GLISS2}
(this gives $R_A\approx 6.37$ and $~5.49$ fm for
Au and Xe, respectively).
{ For Pb nucleus we take the parameters
$R_A=6.62$ and $d=0.546$ fm as in the PHOBOS
Glauber model \cite{PHOBOS,ATDATA}.}

\begin{figure}[!h] 
\begin{center}
\includegraphics[height=5.5cm]{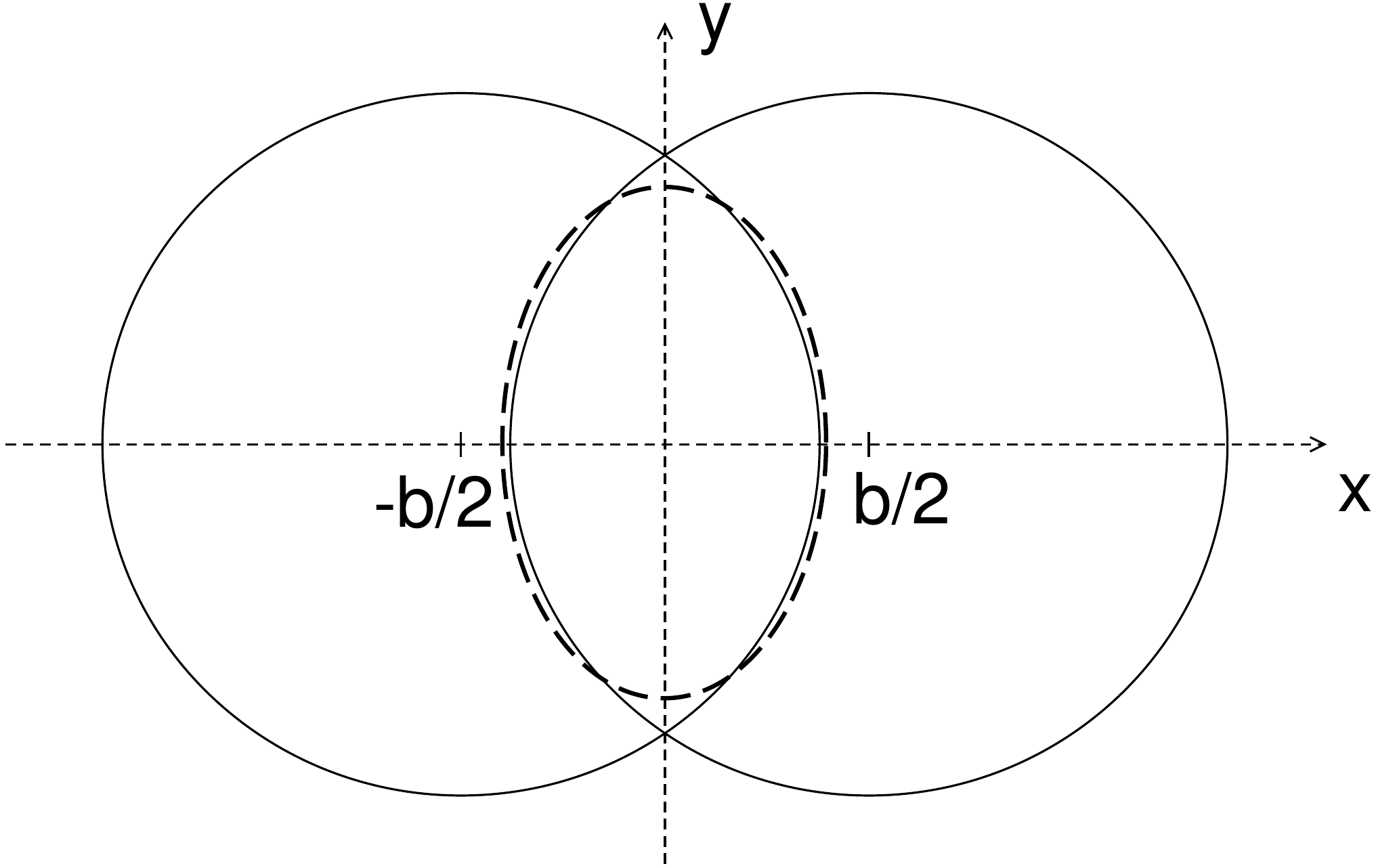}  
\end{center}
\caption[.]{ Schematic representation of the fireball geometry
in  the transverse plane for
  $AA$-collision with impact parameter $b$. See main text for details.
 }
\end{figure}
\begin{figure} 
\begin{center}
\includegraphics[height=5.5cm]{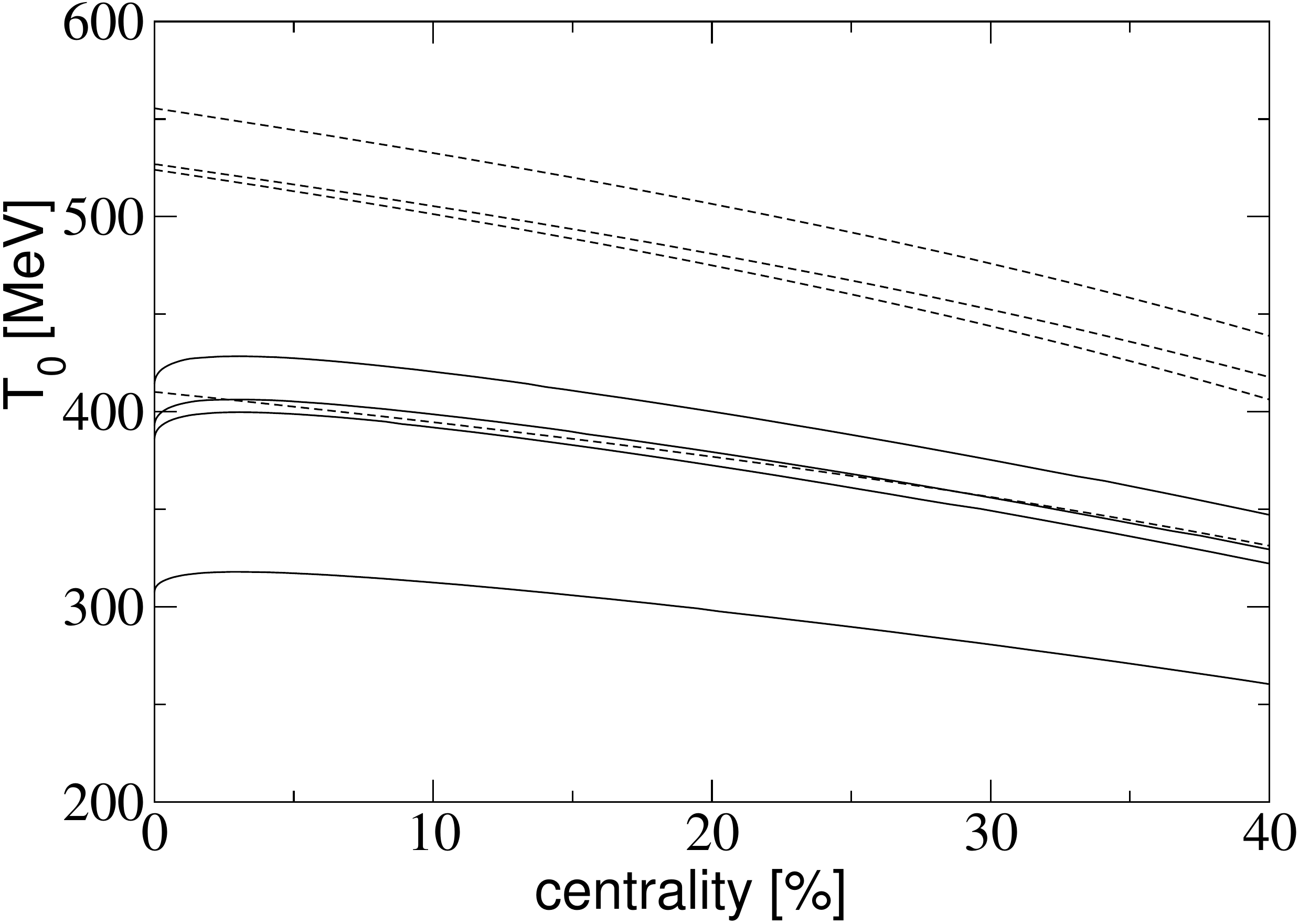}  
\end{center}
\caption[.]{Centrality dependence of the initial QGP temperature obtained
in the Glauber model via the average entropy
  density (solid) and the maximal one at the center of the fireball (dashed)
  for (from bottom to top): $0.2$ TeV Au+Au, $5.44$ TeV Xe+Xe,
  $2.76$ and $5.02$ TeV Pb+Pb collisions. 
 }
\end{figure}
\begin{figure} 
\begin{center}
\includegraphics[height=8.5cm]{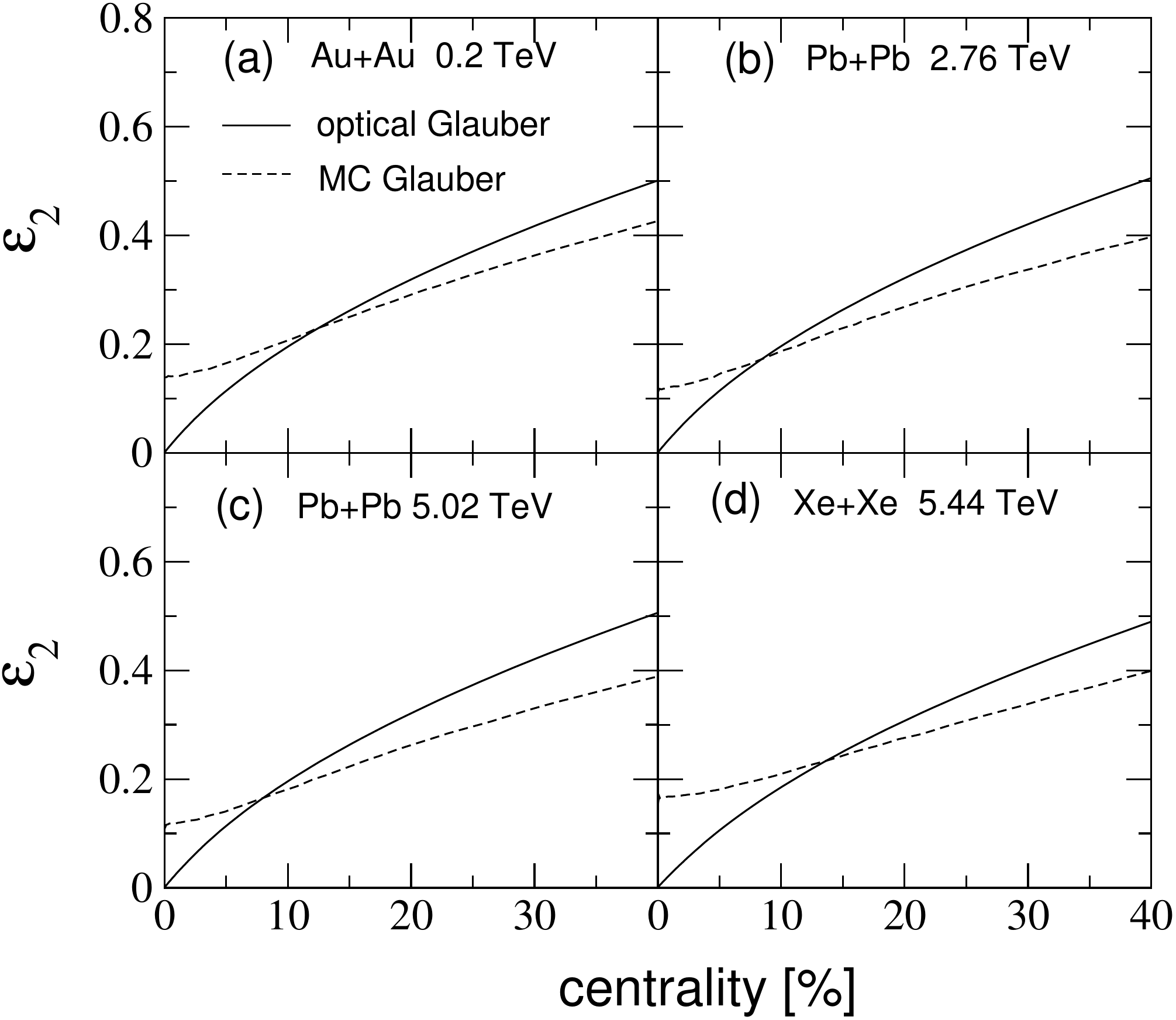}  
\end{center}
\caption[.]{Centrality dependence of the initial fireball eccentricity obtained
in the optical (solid) and Monte-Carlo (dashed) Glauber model
  for: $0.2$ TeV Au+Au collisions (a); $2.76$
  (b) and $5.02$ TeV (c) Pb+Pb collisions;
$5.44$ TeV Xe+Xe collisions (d).
 }
\end{figure}

We calculate both radiative and collisional energy loss with running $\alpha_{s}$ frozen at low momenta at
some value $\alpha_{s}^{fr}$: 
\beq
\alpha_s(Q^2) = \left\{
\begin{array}{ll}
\displaystyle\alpha_{s}^{fr} & \mbox{if } Q \le Q_{fr}\;, \\
\displaystyle\frac{4\pi}{9\log(Q^2/\Lambda_{QCD}^2)} &  \mbox{if } Q > Q_{fr}\;,
\end{array}
\right.
\label{eq:90}
\eeq
where
$Q_{fr}=\Lambda_{QCD}\exp\left\lbrace
{2\pi}/{9\alpha_{s}^{fr}}\right\rbrace$ (in the present analysis we
take $\Lambda_{QCD}=200$ MeV). For gluon emission in vacuum
the value of $\alpha_s^{fr}$ can be estimated from the relation
\beq
\frac{1}{\pi}\int_0^{2~\mbox{GeV}}dQ\alpha_s(Q)\approx 0.38\,,
\label{eq:100}
\eeq
obtained in \cite{DKT} from the heavy quark energy loss.
It gives $\alpha_s^{fr}\approx 1.05$. The constraint on $\alpha_s^{fr}$
in vacuum from (\ref{eq:100}) agrees well with the value of $\alpha_s^{fr}$
obtained from the dipole BFKL analysis of the low-$x$
structure functions \cite{NZ_HERA}. This analysis gives
$\alpha_{s}^{fr}\approx 0.7-0.8$ for $\Lambda_{QCD}=0.3$,
which for $\Lambda_{QCD}=0.2$ corresponds to $\alpha_{s}^{fr}\approx 1 $.   
In the QGP the thermal effects can suppress the in-medium QCD coupling,
and it is reasonable to view $\alpha_s^{fr}$ as a free parameter.
Since we fix the quasiparticle masses $m_{g,q}$, and the temperature
dependence of the Debye mass, $\alpha_s^{fr}$ is  the only free parameter
in our calculations.

\subsection{Model of the QGP fireball}
We perform calculations for the QGP fireball
with purely longitudinal Bjorken's 1+1D expansion \cite{Bjorken}, 
which gives proper time dependence of the entropy density
$s(\tau)/s(\tau_0)=\tau_0/\tau$, where $\tau_0$ is the QGP thermalization
time. Under the assumption that the $n\propto s$,
we have the same $\tau$-dependence for the number density
$n(\tau)=n(\tau_0)\tau_0/\tau$ at $\tau>\tau_0$.
For our basic version we take $\tau_{0}=0.5$ fm. However, to understand
the sensitivity of the results to $\tau_0$ we also performed the calculations
for $\tau_0=0.8$ fm.
For $\tau<\tau_{0}$ we take a linear $\tau$-dependence
$n(\tau)=n(\tau_0)\tau/\tau_0$.
This is just an ad hoc prescription to account for the fact that
the medium production is not an instantaneous process.
As we said, we neglect variation of the initial QGP density with the 
transverse coordinates across the overlapping area of two colliding
nuclei. 
We determine the average initial entropy density of the QGP fireball
from the relation \cite{Bjorken}
\beq
s_{0}=\frac{C}{\tau_{0} S_{f}}\frac{dN_{ch}^{AA}}{d\eta}\,.
\label{eq:110}
\eeq
Here $S_{f}$ is the area of the overlap
region of two colliding nuclei
(as shown in Fig.~1),
and $C=dS/dy{\Big/}dN_{ch}^{AA}/d\eta\approx 7.67$ \cite{BM-entropy} 
is the entropy/multiplicity ratio. 
We define the overlap region as the overlap of two circles with radius
$R=R_{A}+k d$ with $k=2$ ($R_{A}$ and $d$
are the parameters of the Woods-Saxon nuclear density)
\footnote{The value $k\sim 2$ guarantees that for centralities
$\lsim 30$\%, which will be considered in the present analysis,
the fraction of the lost jet cross section is negligible.
Note, that,  in principle, the theoretical  predictions for $R_{AA}$
are not very sensitive to variation of $R$.
It is due to a compensation between the enhancement 
of the energy loss caused by increase of the medium size and its 
suppression caused by reduction of the medium density.}.
We determine the total entropy in the overlap region
using the charged hadron multiplicity pseudorapidity density $dN_{ch}/d\eta$
calculated in the Glauber wounded  nucleon model \cite{KN-Glauber}.
We use the parameters of the model as in our Monte-Carlo
Glauber analyses \cite{Z_MC1,Z_MC2},
which describe very well data on the midrapidity $dN_{ch}/d\eta$
in $0.2$ TeV Au+Au \cite{STAR02_Nch},
$2.76$ \cite{ALICE276_Nch} and $5.02$ TeV \cite{ALICE502_Nch} Pb+Pb, 
and $5.44$ TeV Xe+Xe \cite{ALICE544_Nch} collisions.
For the ideal gas model the entropy density reads
$s(T)=aT^3$ with $a=\frac{4\pi^2}{15}\left(8/3+7N_f/4\right)$
($a\approx 18.53$, if one takes $N_f=2.5$).
In Fig.~2 we show the centrality dependence of initial fireball temperature
for $\tau_0=0.5$ fm for the flat entropy distribution obtained
with the help of the relation (\ref{eq:110}). To illustrate the magnitude of
temperature variation in the transverse plane, we also show there
the Glauber model predictions for the temperature in the center of
the fireball.
Note that the ideal gas approximation
somewhat underestimates the plasma temperature
as compared to that obtained from the lattice entropy density
\cite{t-lat}
(say, $T_{lat}$ is bigger than that  for the ideal gas
by $\sim 5-10$\% at $T\sim 300-450$,
and by $\sim 15-25$\% at $T\sim 150-200$ MeV).
This fact is not important because in our calculations  the ideal
gas temperature plays an auxiliary role 
of a quantity which simply characterizes the entropy density.
The crucial point in our scheme, is the assumption that the
number density is proportional
to the entropy density, and the ratio $n/s$ is
the same as for the ideal QGP (see appendix A). 

Our calculations show that the azimuthally averaged nuclear modification
factor $R_{AA}$ is practically insensitive to the geometry of the fireball. But
this is not the case for the flow coefficient $v_2$, which to good
approximation is proportional to the initial eccentricity $\epsilon_2$
of the fireball. From this point of view, even in
the smooth geometry approximation that we use, the model with
the almond shaped region with two cups turn out to be too crude because
it overestimates $\epsilon_2$ as compared
to that calculated with an accurate transverse dependent entropy density.
For this reason, we
transform the almond shaped interaction region
into an elliptic region of the same area (shown by the dashed line
in Fig.~1).
We will present the results for two variants of the centrality
dependence of $\epsilon_2$ (calculated with the flat
entropy density) for the elliptic overlap region.
For the first variant we 
use the optical Glauber model eccentricity $\epsilon_2$ given by 
\beq
\epsilon_2=\frac{\int dxdy s(x,y)(y^2-x^2)}{\int dxdy s(x,y)(x^2+y^2)}\,.
\label{eq:120}
\eeq
This choice gives $\epsilon_2$ that vanishes 
as centrality tends to zero. In the second variant we use the rms $\epsilon_2$
(it is often denoted $\epsilon_2\{2\}$) obtained in our previous Monte-Carlo
Glauber model simulations \cite{Z_MC1,Z_MC2} of $AA$-collisions.
In this case, due to the density fluctuations, the eccentricity does not
vanishes at zero centrality. Of course, accurate calculations
of the flow coefficients for high-$p_T$ particles require event-by-event
simulations which account for fluctuations of the angle between the
participant plane (that characterizes the orientation
of the fireball ellipse) and the true reaction plane.
In the present analysis we ignore the decorrelation between
these two planes and simply
use the overlap region with fixed orientation (along $y$-axis as shown
in Fig.~1).
The inaccuracy of this approximation is connected with incorrect
treatment of the mutual geometry of the jet production and soft
entropy production. The decorrelation between the geometries of the hard
and soft processes can potentially be important at small centralities.
However, the jet production is concentrated
in the central region of the fireball. For this reason
the inaccuracy of the approximation with the fixed fireball
ellipse orientation should be small because for jet production
at the center of the fireball its orientation
becomes unimportant. We checked this by computing $R_{AA}$ and $v_2$
for different fireball orientations. This test shows that the inaccuracy
due to fluctuations of the participant plane should not be bigger
$5-10$\% for the flow coefficient $v_2$, and negligible for $R_{AA}$.
In Fig.~3 we plot the eccentricity $\epsilon_2$ vs centrality for our two
variants. We will perform comparison with experimental data
for the region $c<30$\%, which corresponds to the impact parameter
$b\lsim 1.1R_A$. We do not consider more peripheral collisions
because for them the approximation of the flat entropy distribution
may become too crude due to enhancement of the boundary effects.
Note that in Fig.~3 $\epsilon_2$ for Xe+Xe collisions
for the Monte-Carlo variant has been obtained accounting for the prolate
shape of the Xe nucleus (see \cite{Z_MC2} for details), which increases
$\epsilon_2$ at $c\sim 0$ by $\sim 15$\%. From Fig.~3 one can see that
the Monte-Carlo $\epsilon_2$ becomes smaller than that for the optical Glauber
model at $c\gsim 10-15$\%. This occurs because fluctuations
of the nucleon positions in the colliding nuclei increase the width of the
almond shaped interaction region when the impact parameter becomes comparable
or larger than the nucleus radius.

The medium life/freeze-out time in $AA$-collisions crucially depends on
the transverse
QGP expansion, which is neglected in the Bjorken model, but becomes very important at $\tau\gsim R_A$ \cite{Bjorken,Broniow_exp}.
The pion interferometry at RHIC \cite{STAR_HBT} and
LHC \cite{ALICE_HBT} energies gives the freeze-out time
$\tau_{f.o.}\approx 1.05\times(dN_{ch}/d\eta)^{1/3}$. We use in our calculations
$\tau_{f.o}$ as the medium life-time, $\tau_{max}$. 
Note that we use the formulas for the parton energy loss in the QGP.
In reality, in the later stage, at $\tau$ close to $\tau_{f.o.}$,
the hot matter is in the hadron gas phase. However, this fact
cannot lead to considerable inaccuracies in the results. The point is
that for a given entropy the transport coefficients in the QGP and hadronic
gas turn out to be close to each other \cite{Baier_q}. And since we
use the QGP number density $\propto s$ our formulas should work in the hadron
gas stage as well. Anyway, in general, the effect of the later stage with
$\tau\sim\tau_{f.o.}$ on the jet modifications is very small.
We checked this by performing calculations also for
$\tau_{max}=0.8\times(dN_{ch}/d\eta)^{1/3}$.

\section{ Numerical results}

\subsection{Optimal $\alpha_s^{fr}$ from the $\chi^2$ fit}
In our previous jet quenching analysis \cite{RAA08,RAA13}
it was found that the LHC data on $R_{AA}$ for
$2.76$ TeV Pb+Pb collisions support somewhat smaller value
of $\alpha_{s}^{fr}$  than the RHIC data for $0.2$ TeV Au+Au collisions.
This conclusion has been made by a simple visual comparison
of the theoretical predictions with the data.
In the present work we perform a more accurate comparison
with data by performing the $\chi^2$-fitting of $R_{AA}$. 
We include only data on $R_{AA}$ because the predictions
of the model for $R_{AA}$ are clearly more robust. In particular the results
for $R_{AA}$ are practically insensitive to the shape of the fireball.
However, as will be seen below our predictions for $v_2$ are in
reasonable agreement with experimental data.
We use the data points for centralities smaller than 30\%.
We include in the $\chi^2$-fitting data
for $0.2$ TeV Au+Au collisions at RHIC \cite{PHENIX_r},
for $2.76$ TeV \cite{ALICE_r276,ATLAS_r276,CMS_r276} and
$5.02$ TeV \cite{ALICE_r502,ATLAS_r502,CMS_r502}
Pb+Pb, and $5.44$ TeV Xe+Xe \cite{ALICE_r544,ATLAS_r544,CMS_r544}
collisions.
 For the lower bound on the particle $p_T$ we take
$p_{T,min}=10$ GeV. { However, for the PHENIX data \cite{PHENIX_r}, which have
a small number of the data points with $p_T>10$ GeV,
we also present the results for $p_{T,min}=7$ GeV.
The value $p_{T,min}\sim 10$ GeV seems to be reasonable from
the point of view of the applicability conditions for our scheme.
Because at lower $p_T$ the leakage of the
probability into the unphysical region $\Delta E_{rad}>E$
(see appendix B) may become too strong. Due to larger energy loss,
this effect is stronger
for gluons. For this reason, the problem  especially concerns the LHC energies,
where at $p_T\lsim 10$ GeV the hadron production comes mostly
from gluon jets}\footnote{{ Note that, without regard to the applicability
  of our approximations at low $p_T$, the inclusion of the data points
 with $p_T\lsim 7-8$ GeV does not make sense, since at such $p_T$
 the non-fragmentation contributions, e.g. from the recombination
 mechanism \cite{Fries,Greco},
  may become important.}}.
For data on $0.2$ TeV Au+Au collisions from PHENIX \cite{PHENIX_r}
we include all data points with $p_T>p_{T,min}$.
For the LHC data we perform the $\chi^2$ analysis  for two versions
of the upper $p_T$-bound: $p_{T,max}=120$ and $20$ GeV.
The latter choice corresponds to the $p_T$-range for the PHENIX
data \cite{PHENIX_r}, and for this reason seems to be preferable
for studying the variation of $\alpha_s^{fr}$ with the QGP density.
For each experiment we calculate $\chi^2$ as
\beq
\chi^2 = \sum_i^N \frac{(f_i^{exp} - f_i^{th})^2}
    {\sigma_{i}^2}\,,
    \label{eq:130}
    \eeq
where $N$ is the number of the data points, the squared errors
include the systematic and statistic errors
$\sigma_i^2=\sigma_{i,stat}^2+\sigma_{i,sys}^2$.

In Fig.~4 we show the variation of $\chi^2/d.p.$
($\chi^2$ per data point) with $\alpha_{s}^{fr}$
in the range from $0.3$ to $1.2$ for different $AA$-collisions and
experiments (the curves have been obtained
for $\tau_{max}=1.05\times(dN_{ch}/d\eta)^{1/3}$) and $\tau_0=0.5$ fm.
In the panel (a) of Fig.~4 we compare the $\chi^2/d.p.$ for the
PHENIX data  for Au+Au collisions
{ (obtained  for $p_{T,min}=7$ and $10$ GeV)} with that
obtained for all the LHC data for $10<p_{T}<120$ and $10<p_{T}<20$ GeV.
In the panels b,~c,~d we plot separately $\chi^2/d.p.$ for
$2.76$ and $5.02$ TeV Pb+Pb, and $5.44$ TeV Xe+Xe
collisions (there we show $\chi^2/d.p.$ for each experiment separately and the
combined $\chi^2/d.p.$).
In calculating $\chi^2$ we have used 
the theoretical $R_{AA}$ obtained with the help of a cubic spline
interpolation for a grid with step $\Delta \alpha_s^{fr}=0.05$
at $0.3<\alpha_s^{fr}<0.5$, and with $\Delta \alpha_s^{fr}=0.1$
at $0.5<\alpha_s^{fr}<1.2$. We also performed the $\chi^2$ fit for
$\tau_0=0.8$ fm. The optimal values of $\alpha_s^{fr}$ and corresponding
values of $\chi^2/d.p.$ for both the versions are summarized in Table I.
For the LHC data we present $\alpha_s^{fr}$ and $\chi^2/d.p.$
separately for each energy (and process) and the combined $\chi^2/d.p.$ for
all LHC experiments. In Table I we also present the results for all
the data sets (RHIC plus LHC).
{
Note that we have checked that the fits with
$p_{T, min}\sim 7-8$ GeV for the LHC data give results very similar to that for
$p_{T, min}=10$ GeV. But the robustness of the results in this case
may be lower than for $p_{T, min}=10$ GeV due to
possibly larger theoretical uncertainties and the contributions
from the non-fragmentation mechanisms \cite{Fries,Greco}. 
  In Table I we present the values of $\alpha_s^{fr}$
  with the standard errors (i.e. corresponding to $\Delta \chi^2=1$). Also,
  we give there the 95\% confidence intervals (CIs) corresponding to
  $\Delta \chi^2$ for the 95\% quantile of the $\chi^2$-distribution.
  The CIs are more appropriate characteristics for understanding
  the difference between the values of  $\alpha_s^{fr}$ for
  RHIC and LHC.

 The results for Pb+Pb collisions are obtained for the Woods-Saxon
 parameters used in the PHOBOS Glauber model \cite{PHOBOS}.
 To understand the sensitivity of the results to the
 choice of the Woods-Saxon parameters we have also performed
 calculations for $R_A=6.49$ and $d=0.54$ as in the GLISSANDO Glauber model
 \cite{GLISS2}. For this set we have obtained the optimal
 $\alpha_s^{fr}$ that are smaller only by $\sim 0.01$.}

\begin{figure}[!h] 
\begin{center}
\includegraphics[height=8.5cm]{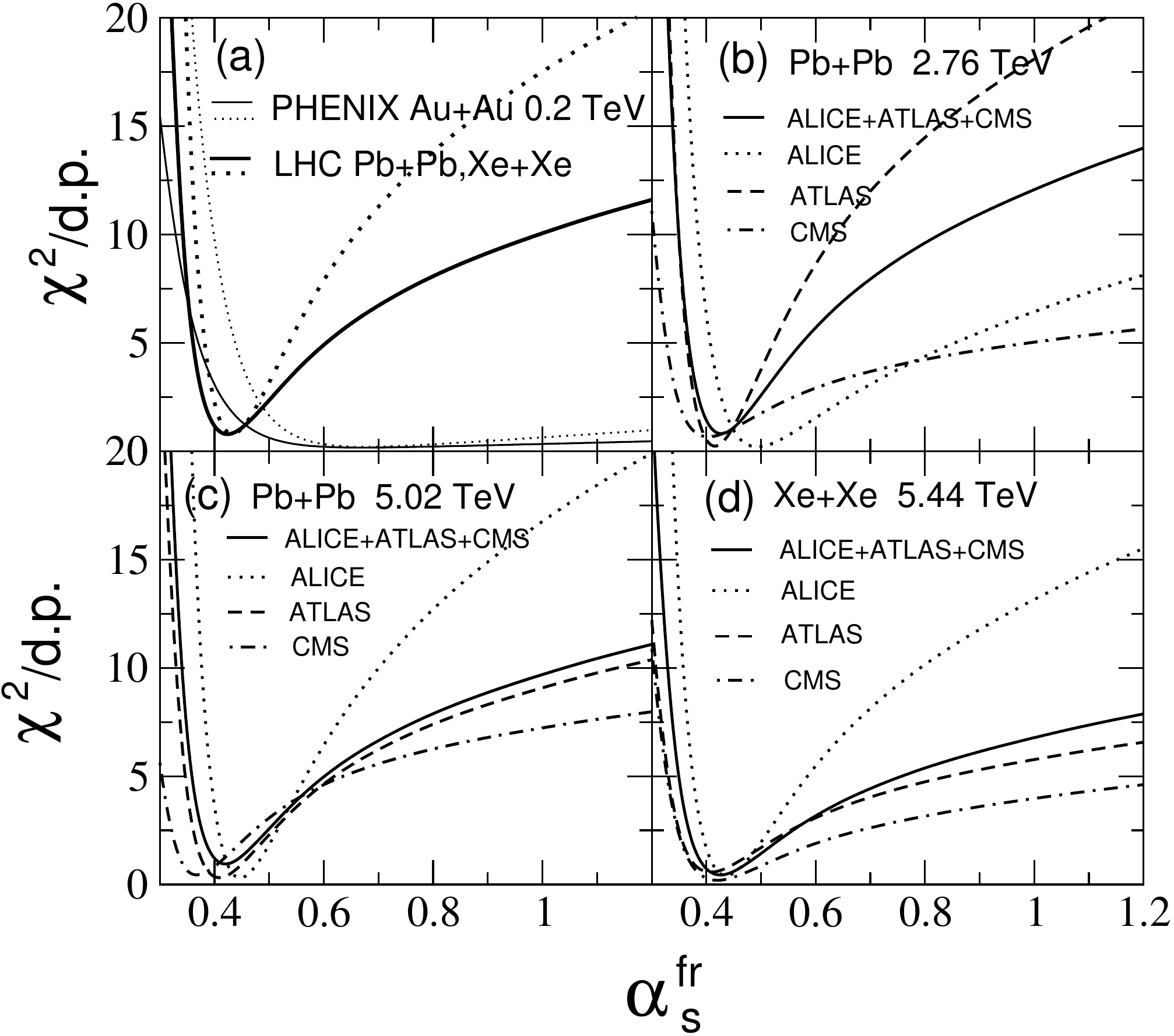}  
\end{center}
\caption[.]
        {The $\chi^2/d.p.$ vs $\alpha_{s}^{fr}$ obtained 
          for $c<30$\% for the version with $\tau_0=0.5$ fm using
          the RHIC and LHC $R_{AA}$ data:
(a) for $\pi^{0}$ at $p_T>10$  GeV (solid line)  and
          $p_T>7$  GeV (dotted line)
in $0.2$ TeV  Au+Au collisions 
from PHENIX \cite{PHENIX_r}  and for $h^{\pm}$
 for $10<p_T<120$ GeV (thick solid line) and for $10<p_T<20$ GeV
(thick dotted line) from the LHC for $2.76$ TeV (from ALICE \cite{ALICE_r276},
ATLAS \cite{ATLAS_r276}, and CMS \cite{CMS_r276})
 and $5.02$ TeV 
 (from ALICE \cite{ALICE_r502},  ATLAS \cite{ATLAS_r502}, and
 CMS \cite{CMS_r502})
Pb+Pb collisions, and
for $5.44$ TeV Xe+Xe collisions 
(from ALICE \cite{ALICE_r544}, ATLAS \cite{ATLAS_r544}) and CMS \cite{CMS_r544};
 (b) for $2.76$ TeV Pb+Pb collisions for the range $10<p_T<120$ GeV;
 (c) same as (b) for  $5.02$ TeV;
 (d) same as (b) for $5.44$ TeV Xe+Xe collisions.
 }
\end{figure}

\begin{table}
  \hspace{-2cm}
         {
           \footnotesize
  \begin{tabular}{c|ccc|ccc}
    \hline
& \multicolumn{3}{c}{$\tau_0=0.5$ fm} &
\multicolumn{3}{|c}{$\tau_0=0.8$ fm} \\
\cline{2-7}
& $\alpha_s^{fr}$ &95\% CI  &$\chi^2/d.p.$  & $\alpha_s^{fr}$ &
95\% CI &  $\chi^2/d.p.$  \\
\hline
Au+Au~0.2~TeV,  $p_T>10$ GeV  &$ 0.67_{-0.08}^{+0.137} $&
$(0.443,>\!\!1.2)$
&$0.167$
&$0.8_{-0.129}^{+0.232}$ &$(0.472,>\!\!1.2)$ & $0.167$\\
Au+Au~0.2~TeV,  $p_T>7$ GeV &$ 0.676_{-0.037}^{+0.046} $& $(0.511,>\!\!1.2)$
&$0.196$&$0.823_{-0.066}^{+0.083}$ &$(0.56,>\!\!1.2)$ & $0.186$\\
\hline
Pb+Pb~2.76~TeV, $10<p_T<120$~GeV &$0.427_{-0.004}^{+0.004}$
&$(0.405,0.456)$ & $0.81$ &
$0.461_{-0.005}^{+0.005}$& $(0.435,0.498)$ & $0.873$\\
Pb+Pb~2.76~TeV, $10<p_T<20$~GeV  &$0.436_{-0.005}^{+0.005}$&$(0.419,0.457)$ &
$0.965$ &$0.475_{-0.006}^{+0.007}$& $(0.453,0.503)$ & $1.016$\\
\hline
Pb+Pb~5.02~TeV, $10<p_T<120$~GeV  &$0.42_{-0.005}^{+0.005}$&$(0.398,0.448)$ & $0.95$ &
$0.456_{-0.006}^{+0.007}$&$(0.432,0.49)$ & $1.021$\\
Pb+Pb~5.02~TeV, $10<p_T<20$~GeV  &$0.43_{-0.006}^{+0.006}$&$(0.412,0.452)$ & $1.01$ &
$0.472_{-0.008}^{+0.009}$&$(0.447,0.506)$ & $1.048$\\
\hline
Xe+Xe~5.44~TeV, $10<p_T<120$~GeV  &$0.42_{-0.006}^{+0.006}$&$(0.382,0.478)$&$0.518$ &
$0.455_{-0.007}^{+0.009}$&$(0.409,0.537)$ & $0.544$\\
Xe+Xe~5.44~TeV, $10<p_T<20$~GeV  &$0.428_{-0.007}^{+0.008}$&$(0.392,0.482)$&$0.267$ &
$0.467_{-0.01}^{+0.011}$&$(0.422,0.548)$ & $0.257$\\
\hline
All LHC, $10<p_T<120$~GeV  &$0.424_{-0.003}^{+0.003}$&$(0.4,0.454)$ & $0.78$ &
$0.459_{-0.004}^{+0.004}$&$(0.432,0.496)$ & $0.836$\\
All LHC, $10<p_T<20$~GeV  &$0.432_{-0.004}^{+0.004}$&$(0.413,0.457)$ & $0.794$ &
$0.473_{-0.005}^{+0.005}$&$(0.447,0.508)$ & $0.805$\\
\hline
RHIC+LHC, $10<p_T<120$~GeV  &$0.426_{-0.003}^{+0.003}$&$(0.405,0.443)$ & $0.874$ &
$0.462_{-0.004}^{+0.004}$&$(0.437,0.497)$ & $0.904$\\
RHIC+LHC, $10<p_T<20$~GeV  &$0.436_{-0.004}^{+0.004}$&$(0.419,0.456)$ & $0.967$ &
$0.477_{-0.005}^{+0.005}$&$(0.453,0.51)$ & $0.924$\\
\hline

  \end{tabular}
}  
  \caption{
    Optimal values of $\alpha_s^{fr}$ with $1\sigma$ standard error,
    95\% CI and corresponding $\chi^2/d.p.$
    obtained for $\tau_0=0.5$ and $0.8$ fm at
    $\tau_{max}=1.05\times(dN_{ch}/d\eta)^{1/3}$
    for different data sets.  For RHIC the $\chi^2$ fits are performed
    for the $R_{AA}$ data points with $p_T>10$ and $7$ GeV.
    For LHC the results are presented
    for $10<p_T<120$ GeV and for $10<p_T<20$ GeV $p_T$-ranges.
\label{chi2fit}}
\end{table}

{ From Table I it is seen that for all the data sets we have a good
  fit quality ($\chi^2/d.p.\lsim 1$).}
From Fig.~4 and Table I one can see that for experimental data from LHC
the optimal values of $\alpha_s^{fr}$ for different energies/processes
turn out to be very similar, but they are noticeably smaller
than the optimal $\alpha_s^{fr}$ for Au+Au collisions at RHIC.
{ The values $\alpha_s^{fr}$ for the joint fits of the RHIC and LHC data
turn out to be close to that for the LHC data. This occurs because
the number of the LHC data points is considerably larger than for the
RHIC data. Note, although the values of $\chi^2/d.p.$ for the combined fits
(RHIC plus LHC) are quite good ($\chi^2/d.p.<1$), the quality of the fits
for the RHIC data taken separately are not good
(for $\tau_0=0.5$ fm $\chi^2/d.p.\approx 1.7$ and $2$ for
$p_{T,max}=20$ and $120$ GeV for the LHC set, respectively).}
For $\tau_0=0.5(0.8)$ fm we have
$\alpha_s^{fr}(\mbox{RHIC})/\alpha_s^{fr}(\mbox{LHC})\approx 1.58(1.74)$ for
the LHC $p_T$-region
$10<p_T<120$ GeV. The situation remains practically the same
for the narrow LHC $p_T$-region (as for the PHENIX
data on Au+Au collisions) $10<p_T<20$ GeV, which gives
$\alpha_s^{fr}(\mbox{RHIC})/\alpha_s^{fr}(\mbox{LHC})\approx 1.55(1.69)$.
{ These ratios will be a bit larger if we use 
  $\alpha_s^{fr}(\mbox{RHIC})$ for the fit with $p_{T,min}=7$
  that gives slightly bigger valiues of $\alpha_s^{fr}$ (see Table I).
  The difference between the optimal values of $\alpha_s^{fr}$ for
  RHIC and LHC is well seen from comparison of the CIs for RHIC and LHC.
  From Table I it is well seen that the the 95\% CIs
  for RHIC and LHC have very small
  overlaps\footnote{{ Note that for the investigate region
$0.3<\alpha_s^{fr}<1.2$ the upper boundaries of the
  95\% CIs for RHIC are not reached. However, this fact
is immaterial from the point of view of the overlap
of the CIs for RHIC and LHC.}}.
  We checked that
  the 68\% CIs for the RHIC and LHC data sets do not overlap at all.
  These facts  say that the difference between the values of
$\alpha_s^{fr}$ for RHIC and LHC is a statistically significant effect. 
}
Thus, similarly to our previous analyses \cite{RAA11,RAA13} with visual
comparison of the theoretical predictions for $R_{AA}$ with data
from RHIC and LHC, the present analysis, with accurate
$\chi^2$-fitting, demonstrates a significant reduction of
the in-medium QCD coupling from RHIC to LHC.

Note that the values of $\chi^2/d.p.$ for $\tau_0=0.5$ and $0.8$ fm
presented in Table I are very similar. This says that jet quenching has
rather weak sensitivity to the medium formation time, and cannot
constrain the value of $\tau_0$. Below we demonstrate this fact
by plotting the predictions for $R_{AA}$ for our two choices of $\tau_0$.
Physically, the fact that jet quenching is weakly sensitive to the initial
stage of the medium evolution is due to strong suppression of the induced
gluon emission by the finite-size effects in the regime
when the the effective gluon formation length is small as
compared to the medium thickness \cite{Z_OA}.
It is worth noting that if the thermalization time decreases with increasing
temperature (say, if $\tau_0\propto 1/T_0$), then it is natural
to calculate the ratio
$\alpha_s^{fr}(\mbox{RHIC})/\alpha_s^{fr}(\mbox{LHC})$
using for RHIC and LHC the values of $\alpha_s^{fr}$ obtained
for $\tau_0=0.8$ and $0.5$ fm, respectively.
In this case for the range $10<p_T<20$ GeV one obtains
$\alpha_s^{fr}(\mbox{RHIC})/\alpha_s^{fr}(\mbox{LHC})\approx 1.85$.

\begin{figure}[!h]
\begin{center}
\includegraphics[height=8.5cm]{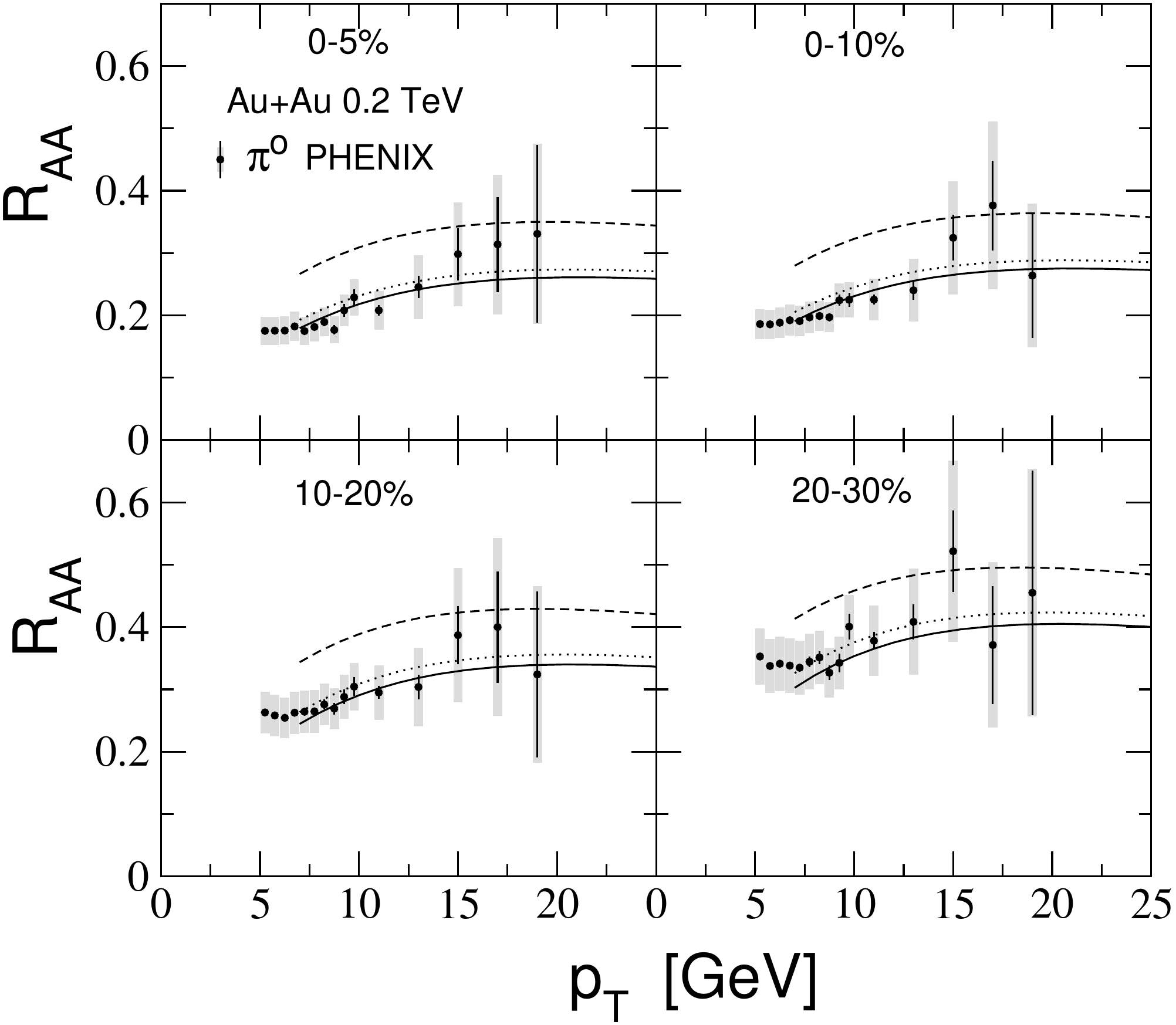}  
\end{center}
\caption[.]{$R_{AA}$ of $\pi^{0}$ for $0.2$ TeV  Au+Au collisions
  from our calculations for $\tau_0=0.5$ (solid) and $0.8$ fm (dotted)
  compared to data from PHENIX \cite{PHENIX_r}.
The solid and dotted curves are for $\alpha_s^{fr}=0.67$ obtained 
by fitting $R_{AA}$ in the range $p_T>10$ GeV
for the version with $\tau_0=0.5$ fm, and the dashed curves are
for $\alpha_s^{fr}=0.424$ obtained 
by fitting the LHC data on $R_{AA}$.}
\end{figure}
\begin{figure}[!h] 
\begin{center}
\includegraphics[height=5.5cm]{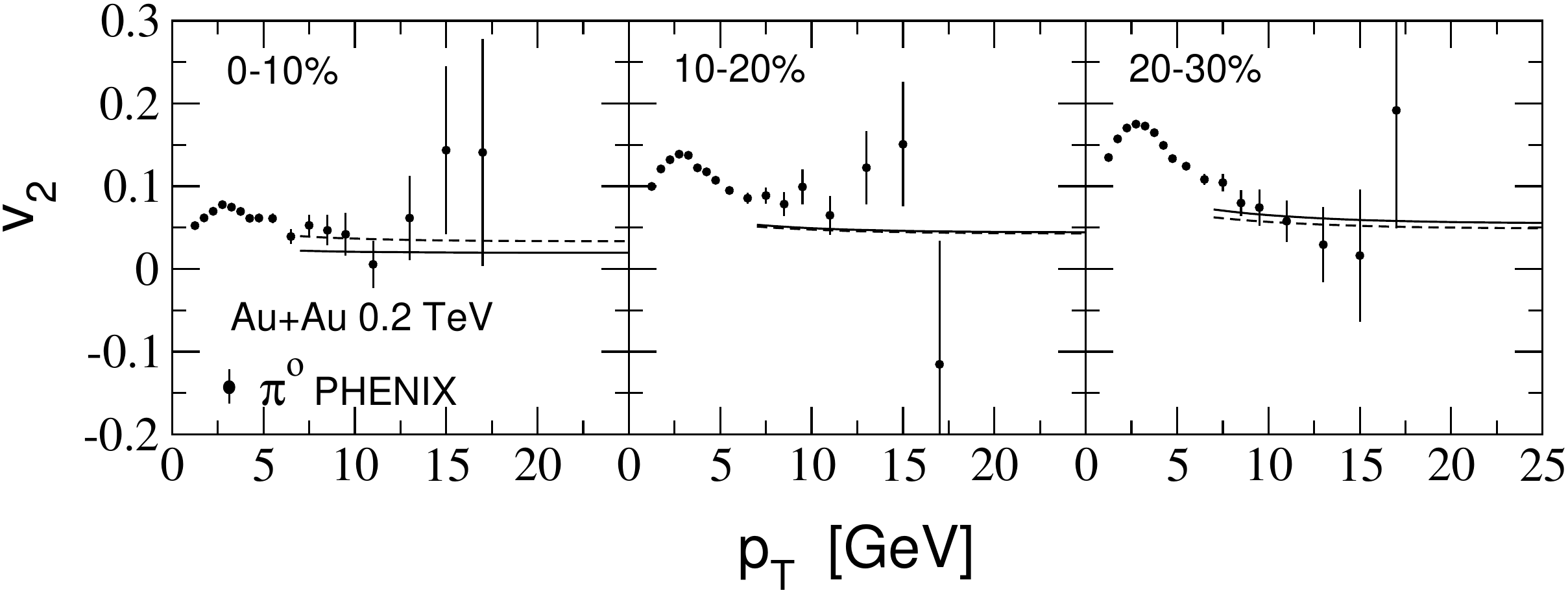}  
\end{center}
\caption[.]
{$v_2$ of $\pi^{0}$ for $0.2$ TeV  Au+Au collisions
obtained for $\tau_0=0.5$ with $\alpha_s^{fr}=0.67$ for
the initial fireball eccentricity $\epsilon_2$ calculated 
in the optical (solid) and Monte-Carlo (dashed) Glauber model (see text
for details). Data points are from 
PHENIX \cite{PHENIX_v}.
}
\end{figure}
\begin{figure} [!h]
\begin{center}
\includegraphics[height=8.5cm]{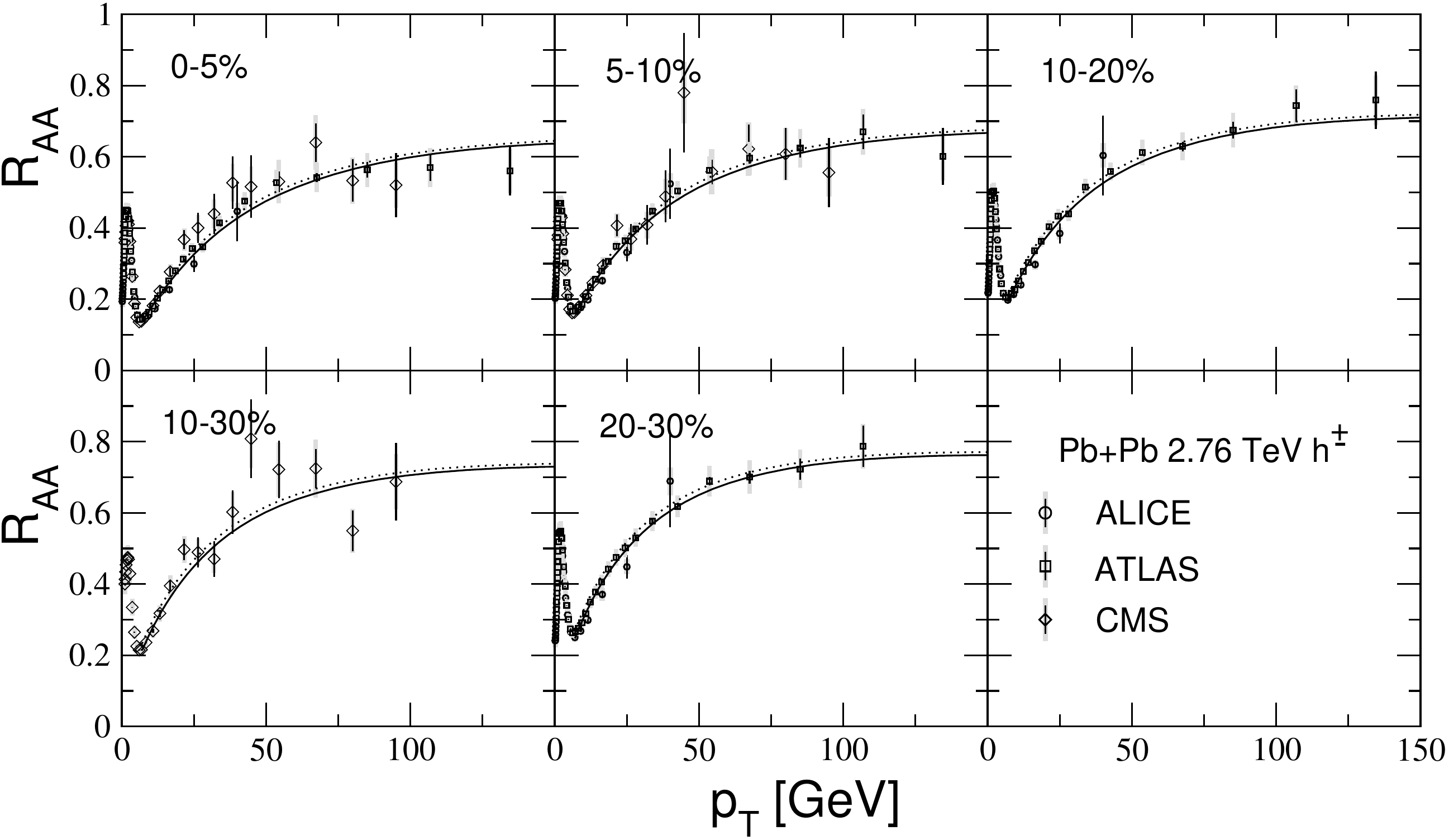}  
\end{center}
\caption[.]
{
$R_{AA}$ of charged hadrons for $2.76$ TeV Pb+Pb collisions
  from our calculations
  for $\tau_0=0.5$ (solid) and $0.8$ fm (dotted)
  with $\alpha_s^{fr}=0.427$ obtained by fitting $R_{AA}$ in the range
$10<p_T<120$ GeV for $\tau_0=0.5$ fm.  
  Data points are from ALICE \cite{ALICE_r276}, ATLAS \cite{ATLAS_r276},
  and CMS \cite{CMS_r276}.
}
\end{figure}
\begin{figure}[!h] 
\begin{center}
\includegraphics[height=8.5cm]{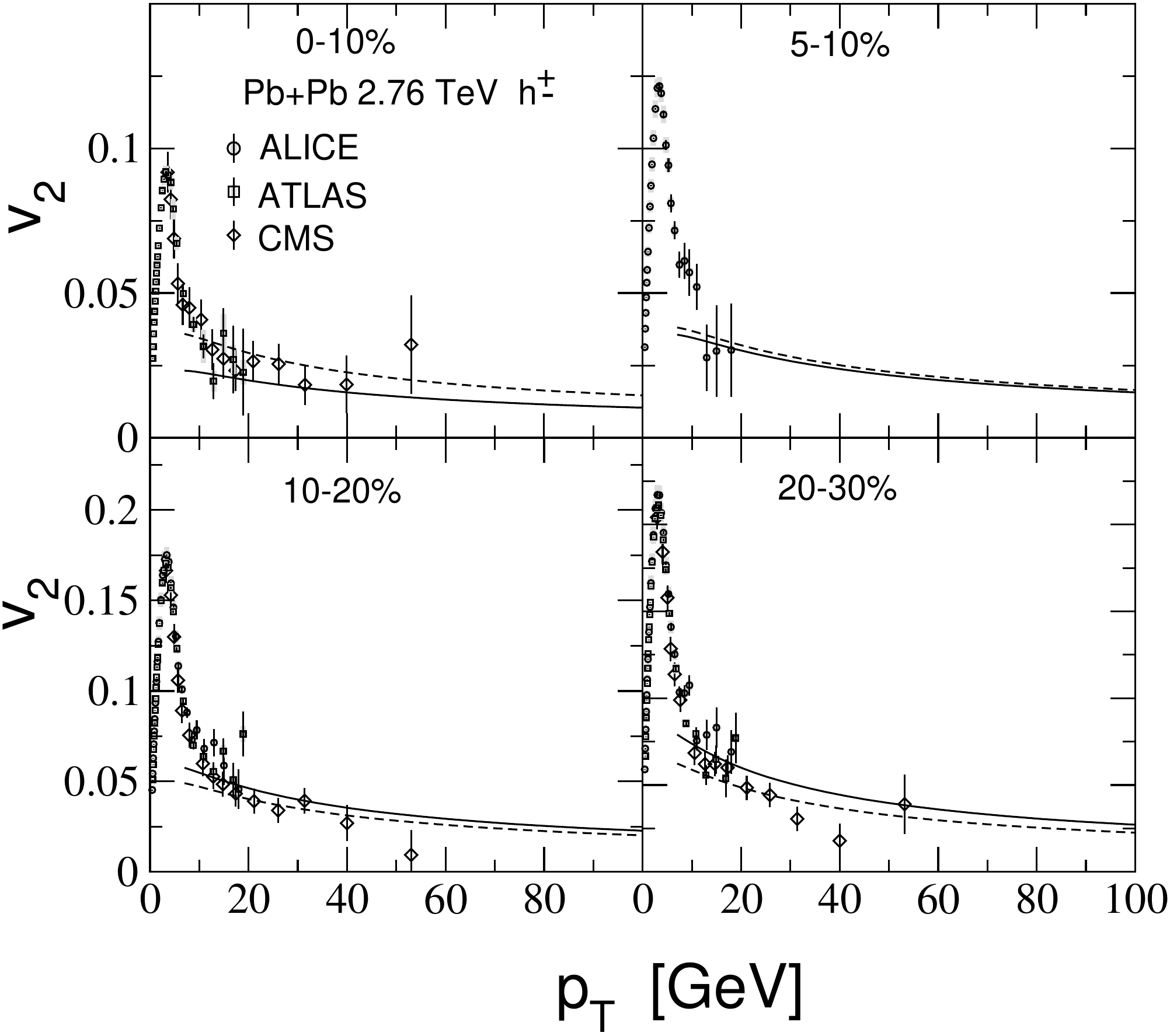}  
\end{center}
\caption[.]{
$v_2$ of charged hadrons in $2.76$ TeV  Pb+Pb collisions
for $\tau_0=0.5$ for the initial fireball eccentricity $\epsilon_2$ calculated 
in the optical (solid) and Monte-Carlo (dashed) Glauber model,
$\alpha_s^{fr}=0.427$ is obtained from the fit of $R_{AA}$ 
in the range
$10<p_T<120$ GeV.
Data points are from ALICE \cite{ALICE_v276}, ATLAS \cite{ATLAS_v276},
  and CMS \cite{CMS_v276}.
}
\end{figure}
\begin{figure} [!h]
\begin{center}
\includegraphics[height=8.5cm]{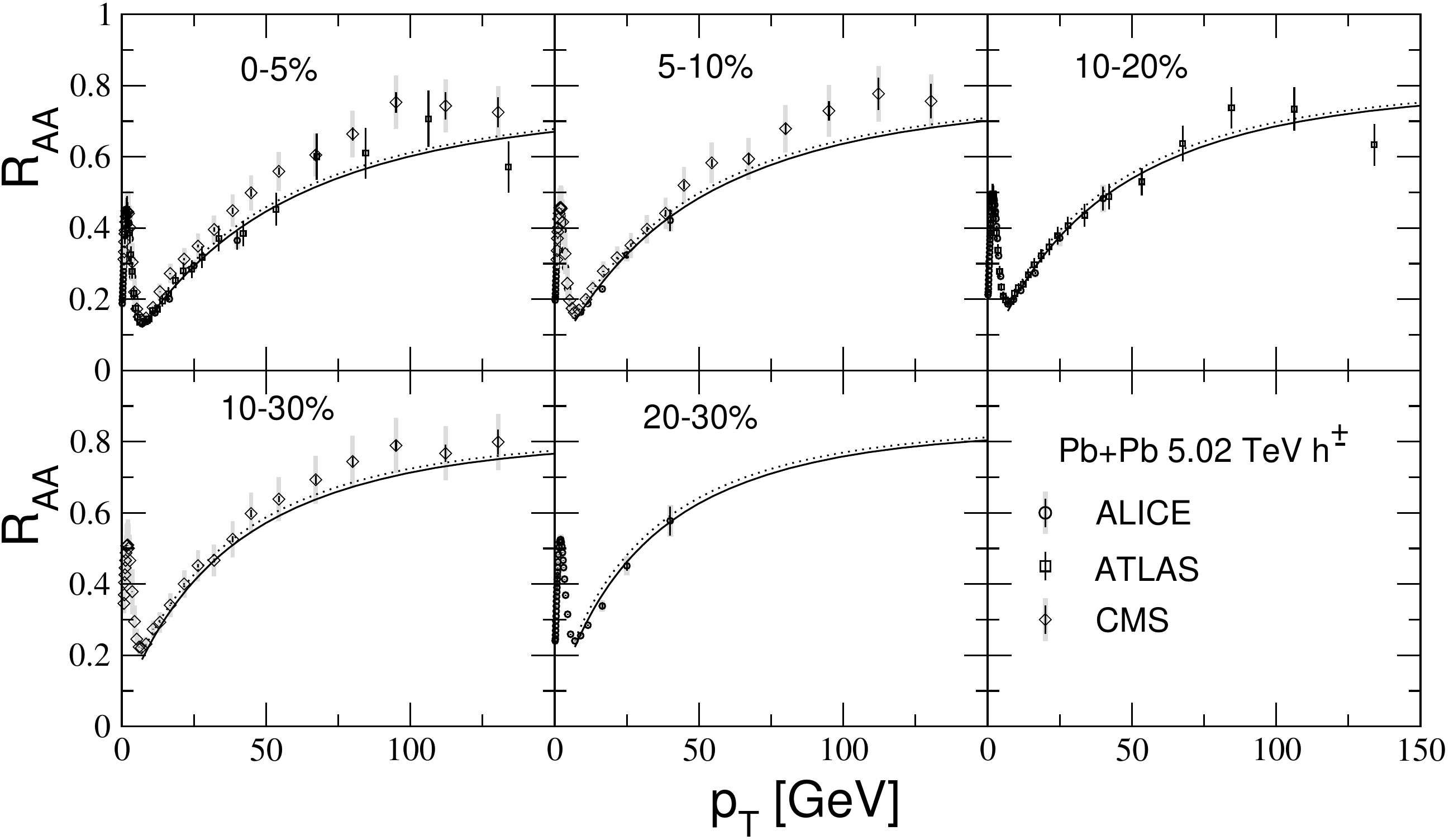}  
\end{center}
\caption[.]
        {Same as in Fig.~7 for $\sqrt{s}=5.02$ TeV for the
          optimal parameter $\alpha_s^{fr}=0.42$.
  Data points are from ALICE \cite{ALICE_r502}, ATLAS \cite{ATLAS_r502},
  and CMS \cite{CMS_r502}.
}
\end{figure}
\begin{figure}[!h] 
\begin{center}
\includegraphics[height=8.5cm]{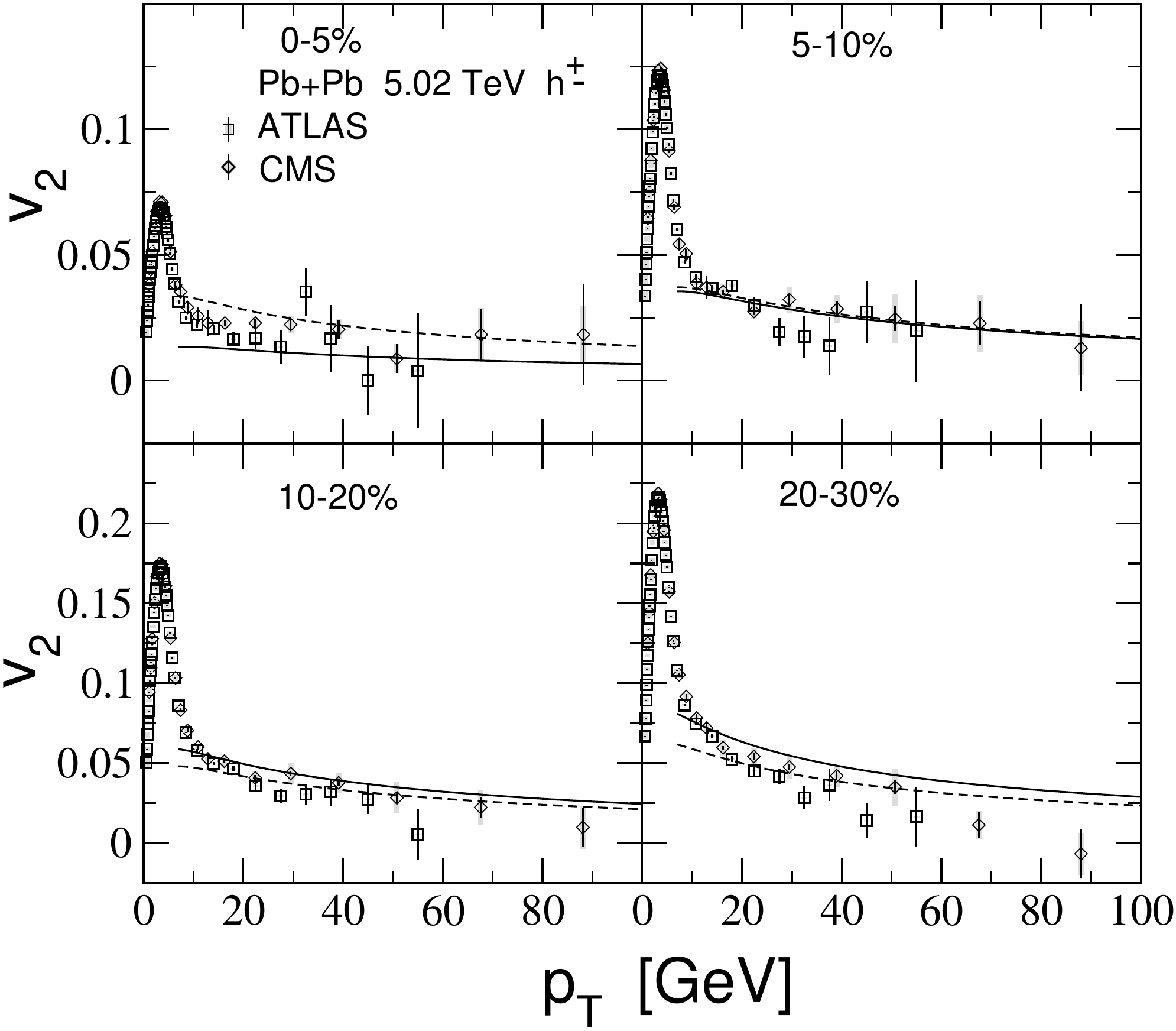}  
\end{center}
\caption[.]
        {Same as in Fig.~8 for $\sqrt{s}=5.02$ TeV
          for $\alpha_s^{fr}=0.42$.
Data points are from ATLAS \cite{ATLAS_v502} and CMS \cite{CMS_v502}.
}
\end{figure}
\begin{figure}[!h]
\begin{center}
\includegraphics[height=8.5cm]{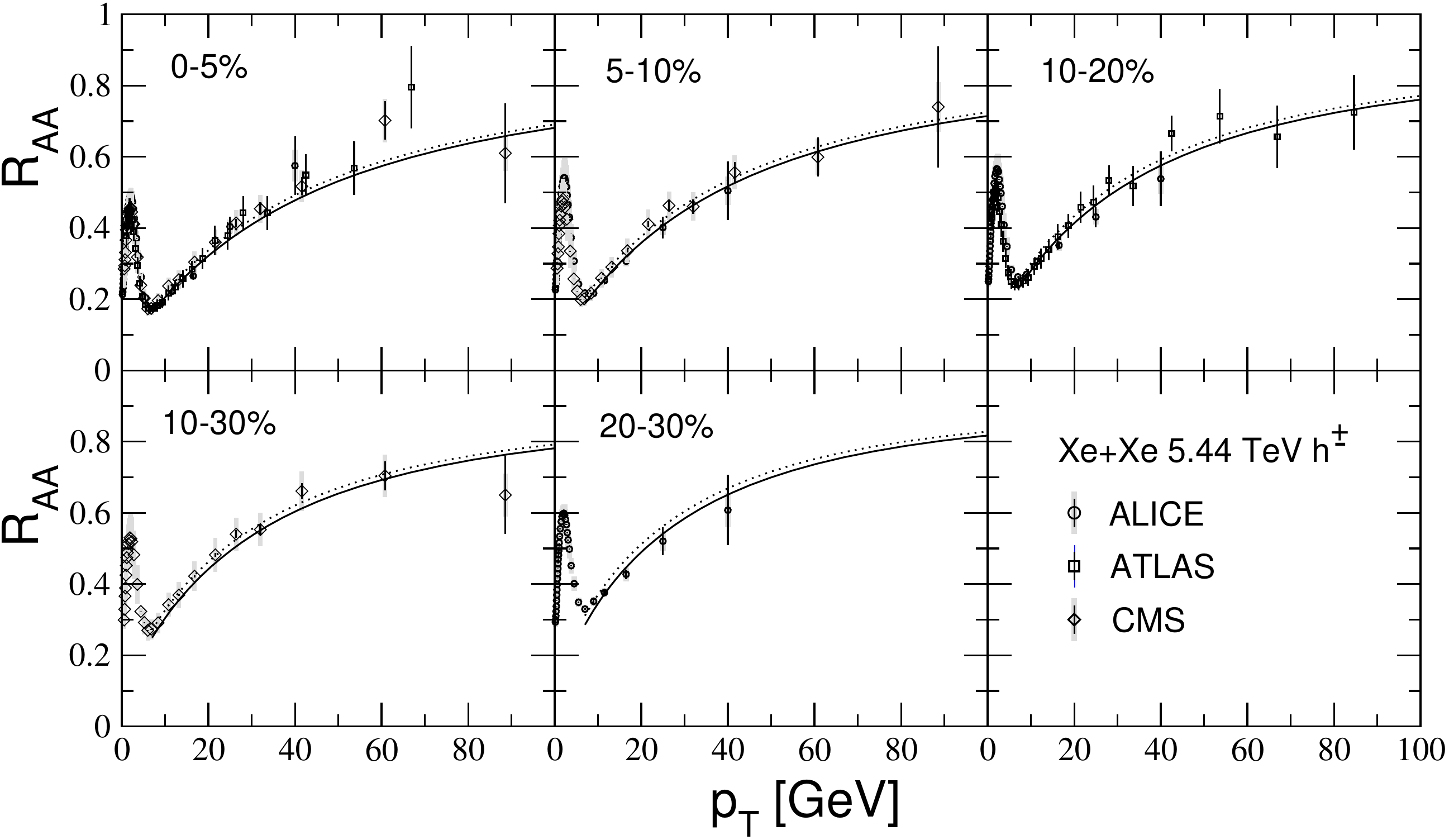}  
\end{center}
\caption[.]
        {
Same as in Fig.~7 for $5.44$ TeV Xe+Xe collisions for the
          optimal parameter $\alpha_s^{fr}=0.425$.
          Data points are from ALICE \cite{ALICE_r544}, ATLAS \cite{ATLAS_r544},
          and CMS \cite{CMS_r544}.
}
\end{figure}
\begin{figure}[!h] 
\begin{center}
\includegraphics[height=8.5cm]{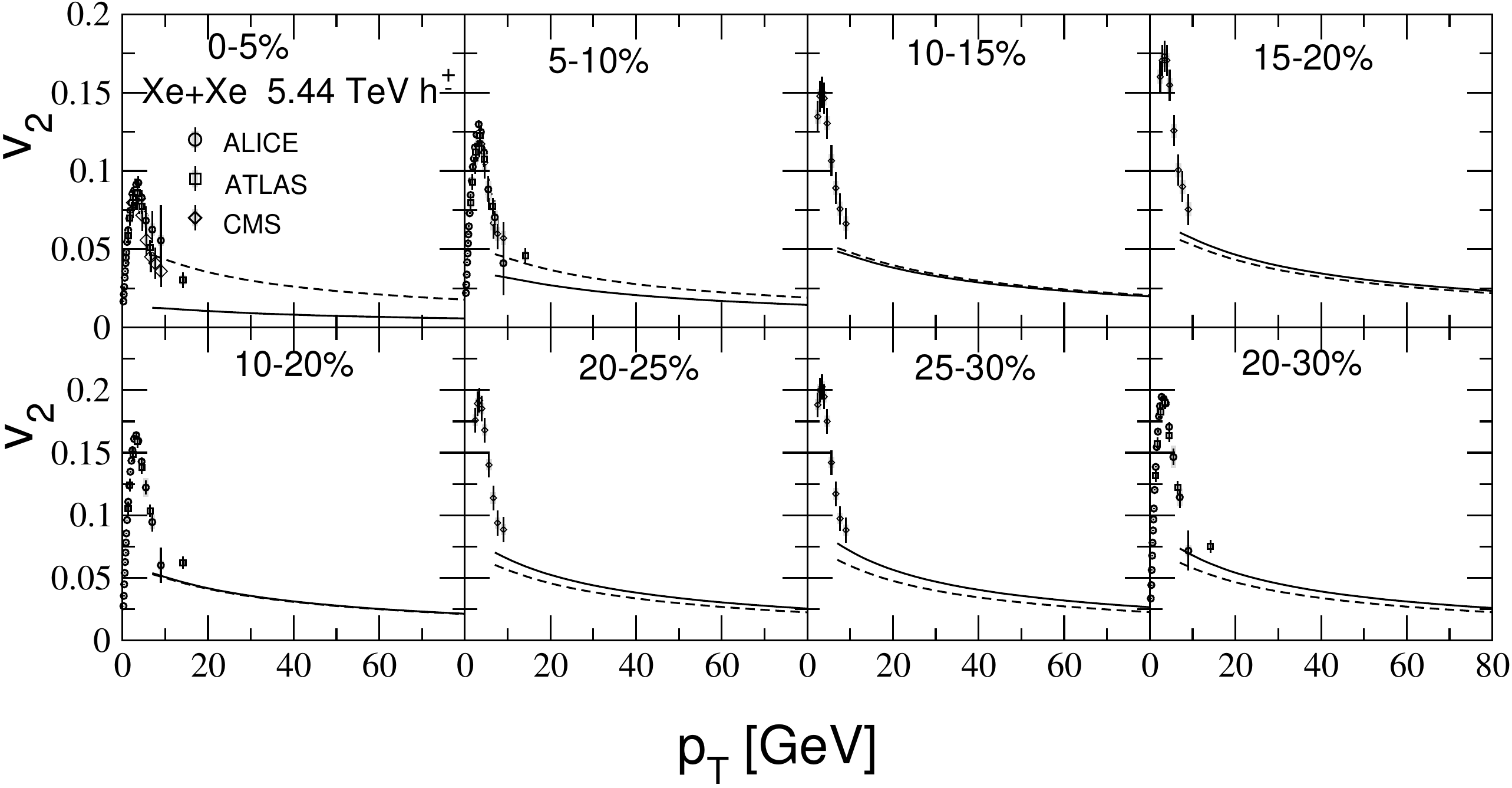}  
\end{center}
\caption[.]
{
Same as in Fig.~8 for $5.44$ TeV Xe+Xe collisions for $\alpha_{s}^{fr}=0.425$.
Data points are from 
ALICE \cite{ALICE_v544}, ATLAS \cite{ATLAS_v544}, and CMS \cite{CMS_v544}.
}
\end{figure}

\subsection{Comparison with experimental data}
In Figs.~5--12 we compare our results for $R_{AA}$ and $v_2$
with data from: PHENIX for $\pi^{0}$-meson in  
$0.2$ TeV Au+Au collisions \cite{PHENIX_r,PHENIX_v};
ALICE \cite{ALICE_r276,ALICE_v276}, ATLAS \cite{ATLAS_r276,ATLAS_v276},
and CMS \cite{CMS_r276,CMS_v276} for $h^{\pm}$ in $2.76$ TeV Pb+Pb collisions; 
ALICE \cite{ALICE_r502}, ATLAS \cite{ATLAS_r502,ATLAS_v502},
and CMS \cite{CMS_r502,CMS_v502} for $h^{\pm}$ in $5.02$ TeV Pb+Pb collisions; 
ALICE \cite{ALICE_r544,ALICE_v544}, ATLAS \cite{ATLAS_r544,ATLAS_v544},
and CMS \cite{CMS_r544,CMS_v544} for $h^{\pm}$ in $5.44$ TeV Xe+Xe collisions.
For each energy/process we show the results
obtained with the optimal value of $\alpha_s^{fr}$ for
$\tau_0=0.5$ fm. To illustrate the sensitivity of $R_{AA}$
to the value of $\tau_0$, we present the results for $\tau_0=0.5$ and $0.8$ fm.
For $R_{AA}$ we show the predictions for the initial eccentricity
$\epsilon_2$ of the fireball obtained in the optical Glauber model.
The theoretical curves for $R_{AA}$ for $\epsilon_2$ obtained in the
Monte-Carlo Glauber model 
are indistinguishable from that for the optical Glauber model version of
$\epsilon_2$. For the flow coefficient $v_2$, where the results
for the two versions of $\epsilon_2$ differ significantly, we
plot the predictions for both the choices of $\epsilon_2$.
{ Note that the results for LHC, shown in Figs.~7--12,
  are very close to that for the optimal $\alpha_s^{fr}$  for the
  combined LHC data set and for the combined RHIC plus
  LHC data set (not shown).}
In Fig.~5, to visualize better the difference between the predictions for
RHIC and LHC, in addition to
predictions for $R_{AA}$ in Au+Au collisions obtained with $\alpha_s^{fr}=0.67$
fitted to the PHENIX data, we also plot the results for
$\alpha_s^{fr}=0.424$ (dashed lines) which is fitted to the LHC data.  As one can see the dashed lines disagree substantially
with the experimental $R_{AA}$. { The
disagreement is especially strong for $p_{T}\lsim 15$ GeV where the
errors become relatively small.
Note that the same situation occurs for the curves for
  the combined RHIC plus LHC fit (not shown), which
  practically coincide with the dashed lines.}

From Figs.~5,~7,~9,~11 one can see that the theoretical predictions for
$R_{AA}$ are in { quite good} agreement with experimental
data\footnote{Note that agreement  with the LHC data on $R_{AA}$ in $2.76$ TeV $0-5$\% Pb+Pb collisions
  is somewhat better than in our previous analysis \cite{RAA13}.  This is mostly due to use in the present
  calculations of $\Lambda_{QCD}=200$ MeV,
  which leads to a bit steeper rise of $R_{AA}$
  at $p_T\gsim 20$ GeV as compared to $\Lambda_{QCD}=300$ MeV
  used in \cite{RAA13}.
  Also, in the present analysis  we use  a physically
  more reasonable algorithm for accounting for the leakage of the
  probability into unphysical region $\Delta E_{rad}>E$ in calculating
  the induced FFs (see appendix B for details). This somewhat
  reduces $R_{AA}$ at $p_T\lsim 15$ GeV for the LHC energies.}. One can see that the difference between our results for $R_{AA}$ obtained with
$\tau_0=0.5$ and $0.8$ fm turns out to be very small (especially for
LHC energies). Note that although we have not included data on the
flow coefficient $v_2$ into our $\chi^2$ analysis, the theoretical
predictions for $v_2$ are in not bad agreement with the data.
Unfortunately, for the PHENIX $v_2$ data \cite{PHENIX_v} large error bars
at $p_T\gsim 10$ GeV render difficult a conclusive comparison, but
within the errors the calculations are consistent with the data.
The situation is better for the LHC measurements of $v_2$
in $2.76$ and $5.02$ TeV Pb+Pb collisions (they give $v_2$ up to
$p_T\sim 50-90$ GeV).
For Pb+Pb collisions we obtain somewhat better agreement with
the LHC $v_2$ data for the Monte-Carlo version of the initial eccentricity
$\epsilon_2$ (see Figs. 8,~9). However, for the optical Glauber model version
the agreement with data is also quite reasonable.
The major difference between the two versions is that the Monte-Carlo
version gives a significantly larger value of $v_2$ for $c\lsim 5$\%.
For 5.44 Xe+Xe collisions
the available data on $v_2$ are restricted to the region $p_T\lsim 14$ GeV, and
a comparison of the theoretical $p_T$-dependence of $v_2$
with experiment is impossible. But nevertheless, from Fig.~12 one
can see that our curves have reasonable matches to the experimental
data points at $p_T\sim 10$ GeV.

The fact that our predictions for $v_2$ are in a  reasonable agreement
with data at $c\sim 20-30$\%
says that it describes correctly the
$L$-dependence of the parton energy loss in the QGP.
Indeed, for such centralities
the typical parton path length in the fireball (see Fig.~1)
for parton momentum along the $y$-axis is bigger than that in the case of
$x$-axis by a factor of $\sim 1.3-1.4$, and to describe the $v_2$
data the model should reproduces correctly the difference in
parton energy losses for these two geometries.
This can also be concluded from description of the difference between $R_{AA}$
for Pb+Pb and Xe+Xe collisions because the Pb nucleus radius is larger
than that for the Xe nucleus by a factor
of $\sim 1.18$.

The curves shown in Figs.~5--12 are obtained for
$\tau_{max}=\tau_{f.o}=1.05\times(dN_{ch}/d\eta)^{1/3}$ \cite{STAR_HBT,ALICE_HBT}.
To check the sensitivity to $\tau_{max}$, we also performed calculations
for $\tau_{max}=0.8\times(dN_{ch}/d\eta)^{1/3}$. We obtained very small difference
between the two versions for the LHC energies. For RHIC the latter version
gives $R_{AA}$ larger by $\sim 3-5$\% at $p_T\sim 10-20$ GeV.
Thus, we see that jet quenching is rather weakly sensitive to the very
initial and
the very late stages of the QCD matter evolution.

Altogether, our calculations show that the pQCD picture
can give a { quite} good agreement with the jet quenching data from RHIC
and LHC.
However, in the present formulation the simultaneous description
of the RHIC and LHC data requires to use different $\alpha_s^{fr}$ at
RHIC and LHC energies. 
A similar difference between jet quenching at RHIC and LHC energies,
in terms of the transport coefficient $\hat{q}$,
has  been found in \cite{JETC_qhat,Salgado_qhat}. 
From the point of view of the QCD matter produced in $AA$-collisions,
the difference in the optimal $\alpha_s^{fr}$/$\hat{q}$ for RHIC and LHC
may be due somewhat stronger thermal suppression
of the effective QCD coupling at the LHC energies.
In order to draw a firm conclusion on this possibility
it is highly desirable to perform calculations with a temperature/density
dependent $\alpha_s$. We leave this to a future analysis.
It is also possible that the bigger values of $\alpha_s^{fr}$/$\hat{q}$
for RHIC mimic an enhancement of parton rescatterings
in the later low temperature stage of the QGP evolution, which should
play a more important role at RHIC energies.
This may be due to presence in the QGP at $T\sim T_c$ \cite{magQGP1}
of the nonperturbative objects like color-magnetic monopoles,
which can enhance the induced gluon emission \cite{Liao_magQGP,Z_magQGP}.
Another cause of the reduction of $\alpha_s^{fr}$ at the LHC energy may
be related to formation of the mini-QGP in $pp$-collisions
\cite{RPP_PRL,RPP14}, which was ignored in the present analysis.
If the mini-QGP formation occurs in $pp$-collisions, the effective
nuclear modification factor turns out to be enhanced by a factor $1/R_{pp}$
(see \cite{RPP14} for details),
where $R_{pp}$ is the modification factor describing jet quenching in the mini-QGP
in $pp$-collisions. Since jet quenching in the mini-QGP should be stronger
at the LHC energies, the effect of the $1/R_{pp}$ factor
should be stronger at the LHC energies. This fact should reduce the
difference between the values of $\alpha_s^{fr}$ for RHIC and LHC.
Note that, if we assume that the mini-QGP is formed only at the LHC
energies, then it is enough to have for LHC $R_{pp}\sim 0.8-0.85$
(such values seems to be very realistic \cite{RPP14})
at $p_T\sim 10-20$ GeV to obtain very similar values of $\alpha_s^{fr}$
for RHIC and LHC.

In principle, the optimal values of  $\alpha_s^{fr}$ for the
RHIC and LHC energies might be affected by the transverse flow effects,
which are neglected in our calculations.
The transverse expansion becomes
very important at later stages ($\tau\gsim R_A$) of the QGP evolution
\cite{Bjorken,Olli_hydro,BF_tflow},
and it is somewhat stronger at the LHC energies.  
But the possibility that
$\alpha_s^{fr}(\mbox{RHIC})/\alpha_s^{fr}(\mbox{LHC})> 1$
is due to different magnitude of the transverse flow at RHIC and LHC
seems to be unrealistic because, as we mentioned above,
the effect of the later stage on $R_{AA}$ is rather small
\footnote{In principle, as was shown in \cite{BMS_tflow} (see also
  \cite{Bass_tflow}), the effect of the radial flow on $R_{AA}$
  is relatively small. This occurs  due to a considerable compensation between
 enhancement of the energy loss caused by increase of the medium size
  and its suppression caused by reduction of the medium density.}.
To understand better whether this scenario is possible, we have performed
calculations using for RHIC the value of 
$\tau_{max}$ corresponding to $2.76$ TeV Pb+Pb collisions (which is bigger
by a factor of $\sim 1.33$).
Even for this unrealistic scenario, we obtained the optimal value
$\alpha_s^{fr}\approx 0.658$ (for $\tau_0=0.5$ fm),
which is very close to that given in table I.
This test shows
that the possibility that
$\alpha_s^{fr}(\mbox{RHIC})/\alpha_s^{fr}(\mbox{LHC})> 1$
is due to ignoring the transverse flow seems to be highly unrealistic.

\section{Conclusions}
In this paper we have performed a detailed comparison
of the pQCD jet quenching calculations
with experimental data on the nuclear modification
factor $R_{AA}$ and the flow coefficient $v_2$ for light hadrons
from RHIC and LHC including the newly available LHC data for
$5.44$ TeV Xe+Xe collisions.
The calculations are performed
within the LCPI \cite{LCPI1} approach using the method suggested
in \cite{RAA04,RAA08}.
We account for radiative and collisional energy loss, 
and fluctuations of the jet path lengths in the QGP. 
The calculations are performed
with running $\alpha_{s}$ frozen at low momenta at
some value $\alpha_{s}^{fr}$, which is treated as a free parameter. 
We have determined the optimal values of $\alpha_{s}^{fr}$ from
the $\chi^2$ fit of $R_{AA}$.
We  have found that for the QGP formation time $\tau_0=0.5$ fm
the RHIC data on $R_{AA}$ in Au+Au collisions
give the optimal value $\alpha_s^{fr}(\mbox{RHIC})\approx 0.67$, while the LHC data
give $\alpha_s^{fr}(\mbox{LHC})\approx 0.42$.
{ The 95\% CIs for $\alpha_s^{fr}(\mbox{RHIC})$ and $\alpha_s^{fr}(\mbox{RHIC})$
have a very narrow overlap region, and the 68\% CIs do not overlap at all.
This clearly shows that the difference between
$\alpha_s^{fr}$ for RHIC and LHC is statistically significant.
For the optimal values of $\alpha_s^{fr}$ our predictions for $R_{AA}$
are in
quite good agreement with experiment. Our $\chi^2$ fitting of $R_{AA}$
for all the data sets gives $\chi^2/d.p.\lsim 1$.
For the flow coefficient $v_2$, which was not included in
the $\chi^2$ analysis, our
predictions are in reasonable agreement with the data as well.}
%
The fact that the model describes well
$R_{AA}$ simultaneously for Pb+Pb
and Xe+Xe collisions, and is in reasonable agreement
with data on  $v_2$ at $c\sim 20-30$\%
says that
the model correctly reproduces the $L$-dependence of the parton energy loss.
Our calculations show that jet quenching
is not very sensitive to the initial and final times of the QGP
evolution.

The difference between the optimal values of $\alpha_s^{fr}$
for RHIC and LHC (which has been also found in our previous
analyses \cite{RAA11,RAA13}) may be due to a somewhat stronger thermal
suppression of the QCD coupling at LHC, or
due to a more important role at RHIC
of the color-monopole states in the QGP
at $T\sim T_c$ \cite{Liao_magQGP}.
Also, this may be related, at least partly,
to the mini-QGP formation in $pp$-collisions, which should affect
differently the predictions for $R_{AA}$ at the RHIC and LHC energies. 
These questions need further investigations.\\
%

I am grateful to O.L.~Kodolova for a helpful communication.
  This work is supported by the Program 0033-2019-0005 of the Russian
  Ministry of Science and Higher Education.

%
\begin{appendix}
\renewcommand{\theequation}{A\arabic{equation}}
\setcounter{equation}{0}

\section*{Appendix A: One gluon spectrum}
In this appendix, we give the formulas used for calculation
of the induced gluon spectrum. We use the representation for the gluon
distribution obtained in \cite{RAA04},
which is convenient for numerical calculations. For 
$q\to g q$ process the gluon spectrum in $x=E_g/E_q$ reads
\beq
\frac{d P}{d
x}=
\int\limits_{0}^{L}\! d z\,
n(z)
\frac{d
\sigma_{eff}^{BH}(x,z)}{dx}\,,
\label{eq:a10}
\eeq
where $n(z)$ is the medium number density, $d\sigma^{BH}_{eff}/dx$ 
is an effective Bethe-Heitler
cross section accounting for both the LPM
and the finite-size effects.
The $d\sigma^{BH}_{eff}/dx$
reads 
\beq
\frac{d
\sigma_{eff}^{BH}(x,z)}{dx}=-\frac{P_{q}^{g}(x)}
{\pi M}\mbox{Im}
\int\limits_{0}^{z} d\xi \alpha_{s}(Q^{2}(\xi))
\left.\frac{\partial }{\partial \rho}
\left(\frac{F(\xi,\rho)}{\sqrt{\rho}}\right)
\right|_{\rho=0}\,\,.
\label{eq:a20}
\eeq
Here 
$P_{q}^{g}(x)=C_{F}[1+(1-x)^{2}]/x$ is the usual splitting
function for $q\to g q$ process,
$
M=Ex(1-x)\,
$
is the reduced "Schr\"odinger mass",
$Q^{2}(\xi)=aM/\xi$ with $a\approx 1.85$ \cite{Z_coll},
$F$ is the solution to the radial Schr\"odinger 
equation for the azimuthal quantum number $m=1$ 
\beq
\hspace{-.2cm} i\frac{\partial F(\xi,\rho)}{\partial \xi}=
\left[-\frac{1}{2M}\left(\frac{\partial}{\partial \rho}\right)^{2}
+v(\rho,x,z-\xi)
+\frac{4m^{2}-1}{8M\rho^{2}}
+\frac{1}{L_{f}}
\right]F(\xi,\rho)\,
\label{eq:a30}
\eeq
with the boundary condition
$F(\xi=0,\rho)=\sqrt{\rho}\sigma_{3}(\rho,x,z)
\epsilon K_{1}(\epsilon \rho)$  
($K_{1}$ is the Bessel function),
$L_{f}=2M/\epsilon^{2}$
with $\epsilon^{2}=m_{q}^{2}x^{2}+m_{g}^{2}(1-x)^{2}$,
$\sigma_{3}(\rho,x,z)$ is the cross section of interaction
of the $q\bar{q}g$ system with a medium constituent
located at $z$.
The potential $v$ in (\ref{eq:a30}) reads
\beq
v(\rho,x,z)=-i\frac{n(z)\sigma_{3}(\rho,x,z)}{2}\,.
\label{eq:a31}
\eeq
The $\sigma_{3}$ is given by
\cite{NZ_sigma3}
\beq
\sigma_{3}(\rho,x,z)=\frac{9}{8}
[\sigma_{q\bar{q}}(\rho,z)+
\sigma_{q\bar{q}}((1-x)\rho,z)]-
\frac{1}{8}\sigma_{q\bar{q}}(x\rho,z)\,,
\label{eq:a40}
\eeq
where
\beq
\sigma_{q\bar{q}}(\rho,z)=C_{T}C_{F}\int d\qb
\alpha_{s}^{2}(q^{2})
\frac{[1-\exp(i\qb\ro)]}{[q^{2}+\mu^{2}_{D}(z)]^{2}}\,
\label{eq:a50}
\eeq
is the local  dipole cross section for the color singlet $q\bar{q}$ pair
($C_{F,T}$ are the color Casimir for the quark and thermal parton 
(quark or gluon), $\mu_{D}$ is the local Debye mass).

For $g\to gg$ one should replace the splitting function and $m_{q}$
by $m_{g}$ in $\epsilon^{2}$. The $\sigma_{3}$ in this case reads
\beq
\sigma_{3}(\rho,x,z)=\frac{9}{8}
[\sigma_{q\bar{q}}(\rho,z)+
\sigma_{q\bar{q}}((1-x)\rho,z)
+\sigma_{q\bar{q}}(x\rho,z)]\,.
\label{eq:a60}
\eeq

As was said in the main text, we assume that the number density of the
QGP is proportional to the entropy density.
Since $\sigma_{q\bar{q}}$ is proportional to the Casimir operator 
of the scattering center, one can treat the QGP as a system of
the triplet color centers with the number density
$n=n_q+n_gC_A/C_F$ (here $n_q$ is the number density of quarks and 
antiquarks, and $n_g$ is the number density of gluons, $C_A$ and $C_F$
are the gluon and quark Casimir operators).
Then, in the ideal gas model,
the effective number density $n$ in the potential (\ref{eq:a31}),
which includes both the quark
and gluons, can be written as $n(z)=bT^{3}(z)$ with
$b=9\xi(3)(N_f+4)/\pi^2\approx 7.125$ (for $N_f=2.5$).

\renewcommand{\theequation}{B\arabic{equation}}
\setcounter{equation}{0}
\section*{Appendix B: Calculation of the induced FFs}
In this appendix, we discuss the method for computation of the induced FFs.
Let us consider first
the case of quark jets. We use the approximation of independent gluon
emission \cite{RAA_BDMS}.
In this approximation the quark distribution in $\xi=\Delta E/E$
in terms of the one gluon spectrum $dP/dx$
can be
written as 
(we omit argument $E$)
\beq
\hspace{-1.9cm}W(\xi)\!=\!W_0\sum_{n=1}^{\infty}\frac{1}{n!}\left[
\prod_{i=1}^{n}\int_0^1 dx_{i}
\frac{dP}{dx_i}
\right]\delta\left(\xi-\sum_{i=1}^{n}x_{i}\right)
\,,\,\,\,\,
W_0=\exp{\left[-\int_{0}^1 dx \frac{dP}{dx}\right]}\,,
\label{eq:b10}
\eeq
where $W_0$ is the no gluon emission probability.
At $\xi\ll 1$ the main effect of the multiple gluon emission
is the Sudakov suppression, which 
reflects a simple fact that
emission of gluons with the fractional momentum bigger than
$\xi$ is forbidden. 
$W(\xi)$ may be written as \cite{GLV2}
\beq
W(\xi)=\sum_{n=1}^{\infty}W_{n}(\xi)\,,
\label{eq:b20}
\eeq
where $W_n$ are determined by the recurrence relations
\bea
W_{n+1}(\xi)=\frac{1}{n+1}
\int_{0}^{\xi}dx W_n(\xi-x)\frac{dP}{dx}\,,\,\,\,\,\,\,\,
W_1(\xi)=W_0\frac{dP}{d\xi}\,.
\label{eq:b30}
\eea
In numerical calculations we set $dP/dx=0$ at $x<m_g/E_q$
and $1-x<m_q/E_q$.

When the average energy loss is small $\langle \Delta E\rangle /E\ll 1$, one
can define the induced $q\to q$ FF
as $D_{q/q}^{in}(z)=W(\xi=1-z)$.
However, for real situation of $AA$-collisions the
ratio $\langle \Delta E\rangle /E$ is not very small, and
the above prescription can violate the
flavor conservation
\beq
\int_0^1 dz D_{q/q}^{in}(z)=1\,
\label{eq:b40}
\eeq
due to a leakage of probability
into the unphysical region of $\xi> 1$ \cite{GLV2,Eskola}.
One can expect that the inaccuracy of the independent gluon emission
picture due to the probability leakage should be concentrated in
the region of $\xi\sim 1$.
At small $\xi$ the $\xi$-dependence of
$W(\xi)$ comes from the Sudakov suppression. Since it is mostly
connected with the one gluon radiation, the approximation of
independent gluon emission should work well for $W(\xi)$ at $\xi\ll 1$.
For this reason to cure the ``flavor nonconservation'' it is
reasonable to modify
somehow $W(\xi)$ only in the region of large $\xi$. In the present analysis
we multiply $W(\xi)$ by a modification factor $K_{qq}$ at $\xi>0.5$,
and determine its value from the flavor conservation (\ref{eq:b40}).

We account for the $q\to g$ transition as well. We define the $q\to g$
FF as
\beq
D_{g/q}^{in}(z)\!=\!K_{gq} dP/dz\,,
\label{eq:b50}
\eeq
where the coefficient $K_{gq}$ is determined from the momentum sum rule
\beq
\int_0^1 dz z\left [D_{q/q}^{in}(z)+D_{g/q}^{in}(z)\right]=1\,.
\label{eq:b60}
\eeq
  
For gluon we account for only $g\to g$ transition.  
We neglect the induced gluon conversion into $q\bar{q}$ pairs,
which for light quarks for RHIC and LHC conditions 
turns out to be relatively small \cite{Z_phot}.
At $z>0.5$, similarly to the $q\to q$ case, we take
$D_{g/g}^{in}(z)=W(\xi=1-z)$, where now $W(\xi)$ is defined via
the one gluon spectrum $g\to gg$.
For $g\to gg$ transition, due to the $x\leftrightarrow  1-x$ symmetry 
of the function $dP/dx$,
we can use $0.5$ for the 
upper limit in $x$-integrations in (\ref{eq:b10})
(we view the softest gluon with $x<0.5$ as a radiated gluon). 
In the soft region $z<0.5$
we take $D_{g/g}(z)=K_{gg}dP/dx$ (with $x=z$). 
We determine the coefficient $K_{g/g}$ from the 
the momentum sum rule
\beq
\int_0^1 dz z D_{g/g}^{in}(z)=1\,,
\label{eq:b70}
\eeq
which should be satisfied (if one neglects the $g\to q\bar{q}$
processes).

Our ansatze on the $z$-dependence of the induced FFs
in the region $z<0.5$ have not serious theoretical motivations.
Fortunately, the form of the induced FFs in the soft region
is practically not important because the typical values of $z$ for
the induced FFs are very close to unity (say, for Au+Au collisions
at $0.2$ TeV $\langle z\rangle\sim 0.9-0.95$ at
$p_T\sim 10-20$ GeV). For this reason the soft region plays a minor role.

Note that the algorithm for calculation of the induced FFs given above
is somewhat different from that used in \cite{RAA11,RAA13}.
In the present analysis, to ensure the probability and momentum
conservation we modify only the FFs in the region $z<0.5$, while
in \cite{RAA11,RAA13} we performed the renormalization
for the whole region of $z$. The latter method leads to somewhat larger
$R_{AA}$ at $p_T\lsim 20$ GeV due an increase of the gluon contribution
(for the quark jets that dominate at RHIC energies the difference
between two methods is small). The new algorithm
seems to be more physically reasonable because, as we said above,
the values of $z$ in the induced FFs that dominate the hadron cross sections
are very close to unity, and this region of $z$ should not be affected
strongly by the leakage of the probability from the region $z\sim 0.5$. 

\end{appendix}

\end{document}